\documentclass[11pt]{article}

\usepackage[12pt]{extsizes}
\usepackage[utf8]{inputenc}
\usepackage{amsmath}
\usepackage{amssymb}
\usepackage{amsthm}
\usepackage{amsfonts}
\usepackage[numbers,sort&compress]{natbib}
\usepackage{graphicx}
\usepackage{lipsum}
\usepackage{enumitem}
\usepackage[font=footnotesize,labelfont=bf]{caption}
\usepackage{mdframed}
\usepackage{authblk}
\usepackage{longtable}
\usepackage{subfigure}
\usepackage[subfigure]{tocloft}
\usepackage[dvipsnames,table]{xcolor}
\usepackage[hidelinks]{hyperref}
\usepackage{mathtools}
\usepackage{relsize}
\usepackage{multirow}
\usepackage{booktabs}
\usepackage{tabularx}
\usepackage{bbding}
\usepackage{listings}
\usepackage{xspace}
\usepackage[ruled,vlined]{algorithm2e}
\usepackage{multirow}
\usepackage[p,osf]{cochineal}
\usepackage[scale=.95,type1]{cabin}
\usepackage[cochineal,bigdelims,cmintegrals,vvarbb]{newtxmath}
\usepackage[zerostyle=c,scaled=.94]{newtxtt}
\usepackage[cal=boondoxo]{mathalfa}
\usepackage{microtype}
\usepackage{multirow}
\usepackage{adjustbox}
\usepackage{etoolbox}
\usepackage{lmodern}
\usepackage{csquotes}
\usepackage[title]{appendix}%

\usepackage{caption}
\usepackage{geometry}
\definecolor{myurlcolor}{HTML}{123463}
\definecolor{dc_color}{RGB}{230, 245, 244}
\definecolor{ds_color}{RGB}{195, 230, 227}
\definecolor{ms_color}{RGB}{150, 214, 209}
\hypersetup{
    colorlinks=true,
    linkcolor=red,
    filecolor=black,      
    urlcolor=black,
    citecolor=blue
}

\apptocmd{\thebibliography}{\raggedright}{}{}

\makeatletter
\patchcmd{\@maketitle}{\LARGE \@title}{\fontsize{30}{19.2}\selectfont\@title}{}{}
\makeatother
\usepackage{cleveref}
\Crefname{section}{Sec.}{Secs.}
\Crefname{equation}{Eq.}{Eqs.}
\Crefname{figure}{Figure}{Figs.}
\Crefname{tabular}{Table}{Tabs.}

\usepackage{tabularx}
\usepackage{ragged2e}
\usepackage{lineno}
\usepackage{chngcntr}
\usepackage{tcolorbox}

\newif\ifhighlighton

\highlightonfalse % to turn off highlight

\ifhighlighton
    \definecolor{customdefaultcolor}{RGB}{18, 159, 87}
    \NewDocumentCommand{\highlight}{O{customdefaultcolor} +m}{%
        \begingroup
        \color{#1}#2%
        \endgroup
    }
\else
    \NewDocumentCommand{\highlight}{O{customdefaultcolor} +m}{#2}
\fi

\usepackage{tocloft}

\usepackage{titletoc}
\newcommand\DoToC{%
  \startcontents
  \printcontents{}{1}{\textbf{Supplementary Materials}\vskip3pt\hrule\vskip5pt}
  \vskip3pt\hrule\vskip5pt
}

\usepackage{minted}
\setminted{
  breaklines,
  breakanywhere,   % allow breaks inside long tokens/paths
  autogobble,      % trim common leading spaces
  fontsize=\scriptsize,
  tabsize=2
}

\usepackage{listings}
\definecolor{bggray}{rgb}{0.97,0.97,0.97}
\definecolor{bordergray}{rgb}{0.85,0.85,0.85}
\definecolor{textgray}{rgb}{0.25,0.25,0.25}
\definecolor{keywordcolor}{rgb}{0.35,0.45,0.85}
\definecolor{commentcolor}{rgb}{0.2,0.6,0.2}
\definecolor{stringcolor}{rgb}{0.65,0.1,0.1}

\usepackage{setspace}
\usepackage{pifont}% http://ctan.org/pkg/pifont

\begin{document}

\newcommand{\method}{DeepEvidence\xspace}

\title{\LARGE \textbf{DeepEvidence: Empowering Biomedical Discovery with Deep Knowledge Graph Research}}

\author[1,\#]{Zifeng Wang}
\author[2]{Zheng Chen}
\author[3]{Ziwei Yang}
\author[2]{Xuan Wang}
\author[4]{Qiao~Jin}
\author[5]{Yifan~Peng}
\author[4]{Zhiyong~Lu}
\author[1,6,\#]{Jimeng~Sun}

\affil[1]{Keiji AI, Seattle, WA, USA}
\affil[2]{Institute of Scientific and Industrial Research, Osaka University, Osaka, Japan}
\affil[3]{Bioinformatics Center, Institute for Chemical Research, Kyoto University, Kyoto, Japan}
\affil[4]{Division of Intramural Research, National Library of Medicine, National Institutes of Health, Bethesda, MD, USA}
\affil[5]{Department of Population Health Sciences, Weill Cornell Medicine, New York, NY, USA}
\affil[6]{School of Computing and Data Science, University of Illinois Urbana-Champaign, Urbana, IL, USA}

\affil[$\#$]{\em{Correspondence: \href{mailto:zifeng@keiji.ai}{zifeng@keiji.ai}; \href{mailto:jimeng@illinois.edu}{jimeng@illinois.edu}}}

\date{}

\maketitle

% \linenumbers


%TC:ignore

\begin{abstract}
Biomedical knowledge graphs (KGs) encode vast, heterogeneous information spanning literature, genes, pathways, drugs, diseases, and clinical trials, but leveraging them collectively for scientific discovery remains difficult. Their structural differences, continual evolution, and limited cross-resource alignment require substantial manual integration, limiting the depth and scale of knowledge exploration. We introduce \method, an AI-agent framework designed to perform Deep Research across various heterogeneous biomedical KGs. Unlike generic Deep Research systems that rely primarily on internet-scale text, \method incorporates specialized knowledge-graph tooling and coordinated exploration strategies to systematically bridge heterogeneous resources. At its core is an orchestrator that directs two complementary agents: Breadth-First ReSearch (BFRS) for broad, multi-graph entity search, and Depth-First ReSearch (DFRS) for multi-hop, evidence-focused reasoning. An internal, incrementally built evidence graph provides a structured record of retrieved entities, relations, and supporting evidence. To operate at scale, \method includes unified interfaces for querying diverse biomedical APIs and an execution sandbox that enables programmatic data retrieval, extraction, and analysis. Across established deep-reasoning benchmarks and four key stages of the biomedical discovery lifecycle: drug discovery, pre-clinical experimentation, clinical trial development, and evidence-based medicine, \method demonstrates substantial gains in systematic exploration and evidence synthesis. These results highlight the potential of knowledge-graph-driven Deep Research to accelerate biomedical discovery.
\end{abstract}

%TC:endignore

% \newpage

\section*{Introduction}

Knowledge graphs have now become foundational to contemporary biomedical discovery, supporting tasks across literature mining, chemical and drug annotation, gene and protein characterization, pathway analysis, and disease-phenotype understanding. Prominent examples include PubTator~\cite{wei2024pubtator} and PKG~\cite{xu2025pubmed} for publication-derived knowledge, KEGG~\cite{kanehisa2017kegg} and Gene Ontology~\cite{gene2019gene} for pathways and functional annotation, OpenTargets~\cite{carvalho2019open} for gene–disease associations, and BioThings~\cite{lelong2022biothings} for unified chemical and gene annotations. Together, these knowledge bases provide rich, complementary representations of biomedical entities, relations, and annotations that underpin modern research workflows. Yet, leveraging them collectively remains highly challenging: they differ markedly in structure, evolve continuously, and rarely align seamlessly, resulting in limited interoperability and cross-resource reasoning difficult in practice. As a result, substantial manual effort is still required to integrate, align, or traverse these heterogeneous knowledge graphs, and existing standardization or merging initiatives remain labor-intensive and difficult to sustain at scale.

Incorporating artificial intelligence (AI) agents into biomedical research has become an area of rapidly growing interest, with applications emerging across systematic literature review~\cite{wang2025foundation,wang2025accelerating}, genomics research~\cite{wang2025making,wang2025geneagent}, spatial biology~\cite{wang2025spatialagent}, biomarker discovery~\cite{swanson2025virtual}, and beyond. Central to this trend is the concept of the AI scientist: LLM-driven agents that interact with diverse scientific knowledge sources and computational tools to streamline, enhance, or automate components of the research process~\cite{gao2024empowering,gottweis2025aicoscientist,boiko2023autonomous}. Alongside these efforts, several general-purpose biomedical agents have also been developed~\cite{huang2025biomni,gao2025democratizing}. Within this landscape, ``Deep Research'' represents a class of LLM-based agents capable of end-to-end scientific inquiry by autonomously searching, reasoning, and synthesizing evidence from literature and knowledge bases, as exemplified by OpenAI DeepResearch~\cite{openaideepresearch}, Gemini Deep Research~\cite{geminideepresearch}, Perplexity Deep Research~\cite{perplexitydeepresearch}, and so on. Despite these advances, substantial challenges remain. In particular, most existing agents rely solely on internet-scale text corpora, leaving open a central question:\textit{ Can we design agents that deeply and systematically explore large, heterogeneous biomedical knowledge graphs, bridging them to uncover the subtle relationships and connections essential for modern biomedical discovery?}

Here, we present \method, an AI agent system specifically designed to enable deep research with biomedical knowledge graphs. Unlike generic Deep Research systems, \method leverages specialized knowledge-graph tooling to locate and link entities across heterogeneous graphs, and to navigate them through a coordinated combination of Breadth-First ReSearch (BFRS) and Depth-First ReSearch (DFRS): two complementary strategies for systematic graph exploration. \method further incorporates an internal evidence-graph memory that is continuously updated as the agent traverses and synthesizes information, providing a structured and transparent record of the discovery process. We instantiate these principles in an orchestrator–sub-agent architecture comprising dedicated BFRS and DFRS agents that collaborate to build and refine the final evidence graph (Figure~\hyperref[fig:figure1]{1}).

Our empirical evaluation begins by assessing \method’s performance across established biomedical benchmarks that require deep exploration and reasoning over literature, including Humanity’s Last Exam (HLE)~\cite{phan2025humanity}, LabBench-LitQA2~\cite{laurent2024lab}, SuperGPQA~\cite{du2025supergpqa}, and TrialPanorama-EvidenceQA~\cite{wang2025trialpanorama}. We then demonstrate \method’s practical utility across four critical stages of the biomedical discovery lifecycle: (1) drug discovery, (2) pre-clinical in vivo experimentation, (3) clinical trial development, and (4) evidence-based medicine. Together, these results illustrate the potential of knowledge-graph-driven Deep Research to support and accelerate biomedical discovery.
\section*{Results}

\subsection*{Overview of \method}
\begin{figure}
    \centering
    \includegraphics[width=0.95\linewidth]{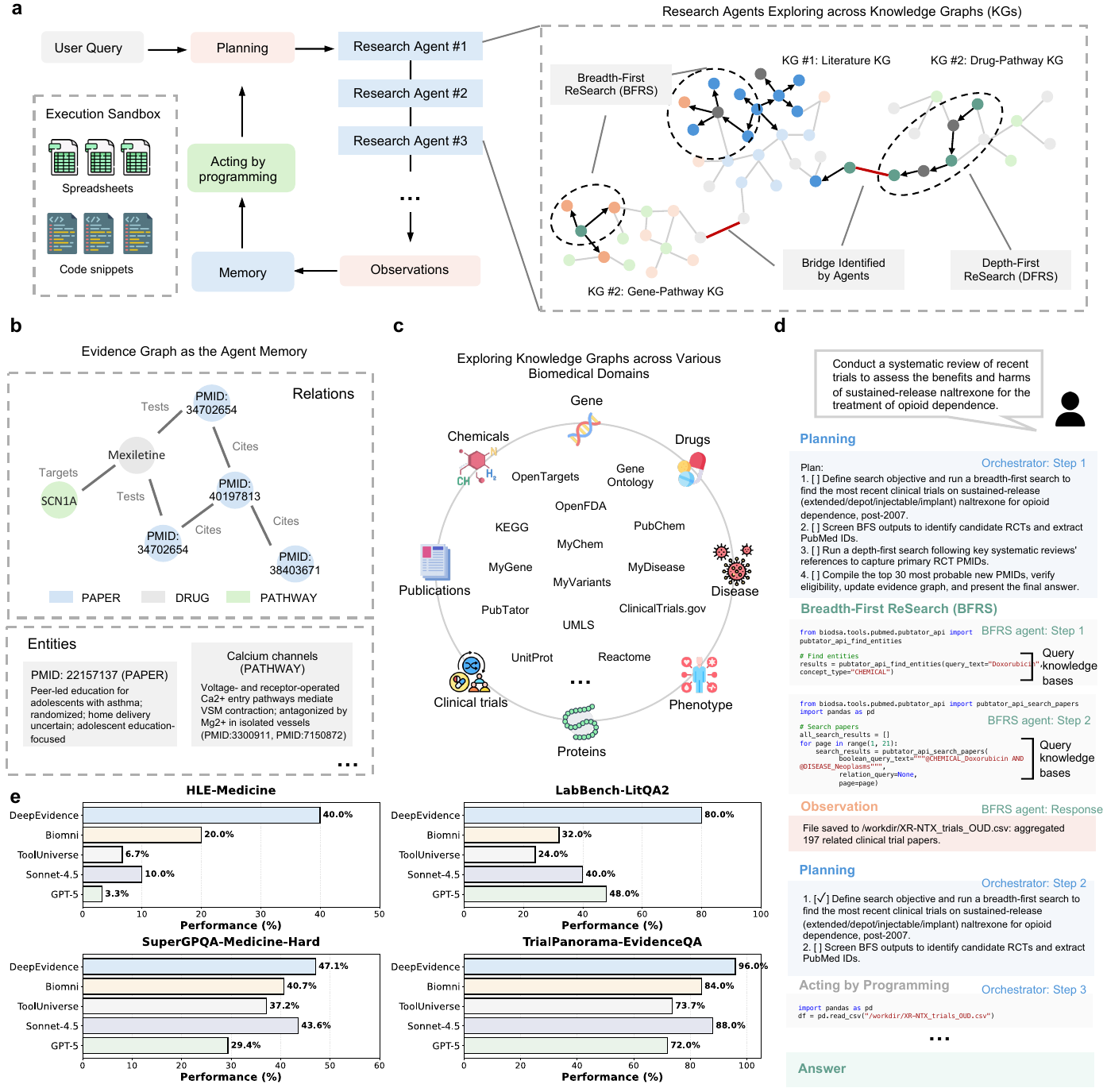}
    \caption{Overview of the \method and the benchmark performance. (a) System overview showing the orchestrator coordinating planning, action, and memory updates across research agents. A user query is translated into a planned sequence of research agents that act within an execution sandbox and update a shared memory through observations. These agents perform coordinated breadth-first and depth-first exploration across multiple biomedical knowledge graphs to identify cross-graph links and build an integrated evidence view.
(b) The agent incrementally builds a structured graph that links papers, drugs, genes, and pathways through relations such as targets, tests, and citations. This evidence graph serves as a persistent memory that organizes discovered entities and supports downstream reasoning.
(c) Overview of the diverse biomedical knowledge graphs that the agent can access across domains, including genes, diseases, drugs, chemicals, proteins, phenotypes, publications, and clinical trials.
(d) A step-by-step example showing how the agent plans, queries knowledge bases, executes breadth-first research, collects observations, and uses programming actions to complete a systematic review task.
(e) Benchmark results showing that \method consistently outperforms strong baseline systems across four biomedical reasoning tasks.}
    \label{fig:figure1}
\end{figure}

\method is a Deep Research framework designed to operate over heterogeneous biomedical knowledge graphs (KGs) with LLM-driven AI agents. At its core is an orchestrator agent that delegates tasks to a set of specialized sub-agents, termed research agents, which execute knowledge-graph exploration (Figure~\hyperref[fig:figure1]{1a}). Throughout the process, the orchestrator maintains and incrementally updates an internal evidence graph that records the key entities and relations critical to answer the query (Figure~\hyperref[fig:figure1]{1b}). The research agents operate in two complementary modes. Breadth-First ReSearch (BFRS) issues broad, multi-graph queries to rapidly survey first-hop neighborhoods (e.g., identifying drugs linked to a disease across disease–drug or gene–disease graphs) and capture a wide set of candidate entities and relations. Depth-First ReSearch (DFRS) builds on the important entities surfaced by BFRS and performs deeper, multi-hop exploration when the question requires multi-step reasoning or richer evidence discovery (e.g., recursively tracing references across publications to construct a comprehensive evidence graph for systematic literature review). We refer to these two modes as BFRS and DFRS, rather than the conventional BFS/DFS, because the proposed AI agents choose traversal steps autonomously rather than follow fixed graph-theoretic rules. Also, AI agents can bridge across heterogeneous knowledge graphs by recognizing shared entities or inferring cross-graph relations (referred to as ``Bridge entities'' in Figure~\hyperref[fig:figure1]{1a}). 

We created an integrated environment and toolset that allow \method to explore diverse biomedical KGs at scale (Figure~\hyperref[fig:figure1]{\ref*{fig:figure1}c}). Many biomedical KGs are accessible through heterogeneous web APIs, so to support efficient BFRS, we implemented unified entity-search and data-fetching interfaces across major biomedical domains, including publications, clinical trials, chemicals, genes, drugs, diseases, phenotypes, and proteins. For example, a single drug query issued by an agent can automatically trigger parallel lookups across KEGG~\cite{kanehisa2017kegg}, ChEMBL~\cite{zdrazil2024chembl}, OpenFDA~\cite{openfdadrug}, OpenTargets~\cite{carvalho2019open}, and other resources. Because these operations involve large numbers of API calls and produce heterogeneous data, we equipped the system with an execution sandbox that enables agents to conduct research programmatically. Within this sandbox, agents can write and execute code to query knowledge-graph APIs at scale, persist retrieved results as structured files (e.g., spreadsheets), and perform downstream extraction and analysis all within a controlled environment. This design ensures that multi-step research workflows remain both efficient and reproducible.

\method is designed to support general biomedical research tasks through an agentic workflow, in which the orchestrator dynamically determines the research plan based on the query (Figure~\hyperref[fig:figure1]{\ref*{fig:figure1}d}). At the outset, the orchestrator selects the sequence and interplay of BFRS and DFRS and identifies the research targets for each step. Each research agent then autonomously determines its own actions, such as which KGs to query and what programs to execute for data extraction and analysis. Nonetheless, the system remains fully steerable: users can specify the biomedical domains relevant to a particular query, thereby constraining the search space, or provide expert priors to guide and accelerate the Deep Research process.

\subsection*{DeepEvidence excels on biomedical research benchmarks}
We evaluated \method on four challenging open benchmarks that require general biomedical knowledge, multi-hop reasoning, and evidence-based research: Humanity's Last Exam (HLE)~\cite{phan2025humanity}, LabBench~\cite{laurent2024lab}, SuperGPQA~\cite{du2025supergpqa}, and TrialPanorama~\cite{wang2025trialpanorama} (Figure 1e). To highlight the contribution of our proposed Deep Research framework, we compared \method against strong baselines: (1) two frontier LLMs (GPT-5 and Sonnet-4.5) equipped with literature-search capabilities, and (2) two general-purpose biomedical agents (Biomni \cite{huang2025biomni} and ToolUniverse \cite{gao2025democratizing}). For fairness, all agent systems, including \method, were run with the same underlying LLM (GPT-5).

Across all four benchmarks, \method consistently delivered the strongest performance (Figure~\hyperref[fig:figure1]{\ref*{fig:figure1}e}). On HLE-Medicine, \method achieved 40.0\%, substantially outperforming Biomni (20.0\%), ToolUniverse (10.0\%), Sonnet-4.5 (3.3\%), and GPT-5 (3.3\%). On LabBench-LitQA2, a benchmark requiring retrieval and synthesis of publication-derived evidence, \method reached 80.0\%, more than doubling Biomni (32.0\%) and surpassing Sonnet-4.5 (48.0\%). On SuperGPQA-Medicine-Hard, which emphasizes difficult mechanistic reasoning, \method achieved 47.1\%, exceeding all baselines (Biomni 40.7\%; Sonnet-4.5 43.6\%). Finally, on TrialPanorama-EvidenceQA, a clinical-evidence reasoning task grounded in real-world clinical trials, \method obtained 96.0\%, outperforming both Biomni (84.0\%) and Sonnet-4.5 (88.0\%).

These results demonstrate the value of deep knowledge-graph research in biomedicine. Although agents like Biomni and ToolUniverse have access to many KG APIs, they typically perform shallow exploration, issuing only a few searches before answering. Their large tool suites can also introduce tool confusion, leading agents to select irrelevant tools or follow incorrect reasoning paths. In contrast, \method carries out structured BFRS–DFRS exploration guided by a continually updated evidence graph, enabling deeper, more targeted, and more reliable KG investigation. This design directly supports superior performance on tasks that require multi-hop reasoning and evidence synthesis.

To further assess \method’s generalization in realistic biomedical research settings, we curated six new benchmark tasks that span the full lifecycle of biomedical discovery, including drug discovery, preclinical research, clinical trial development, and evidence-based medicine. We introduce and analyze each benchmark in detail in the following sections.

\subsection*{DeepEvidence for drug discovery and preclinical research}
\begin{figure}
    \centering
    \includegraphics[width=0.95\linewidth]{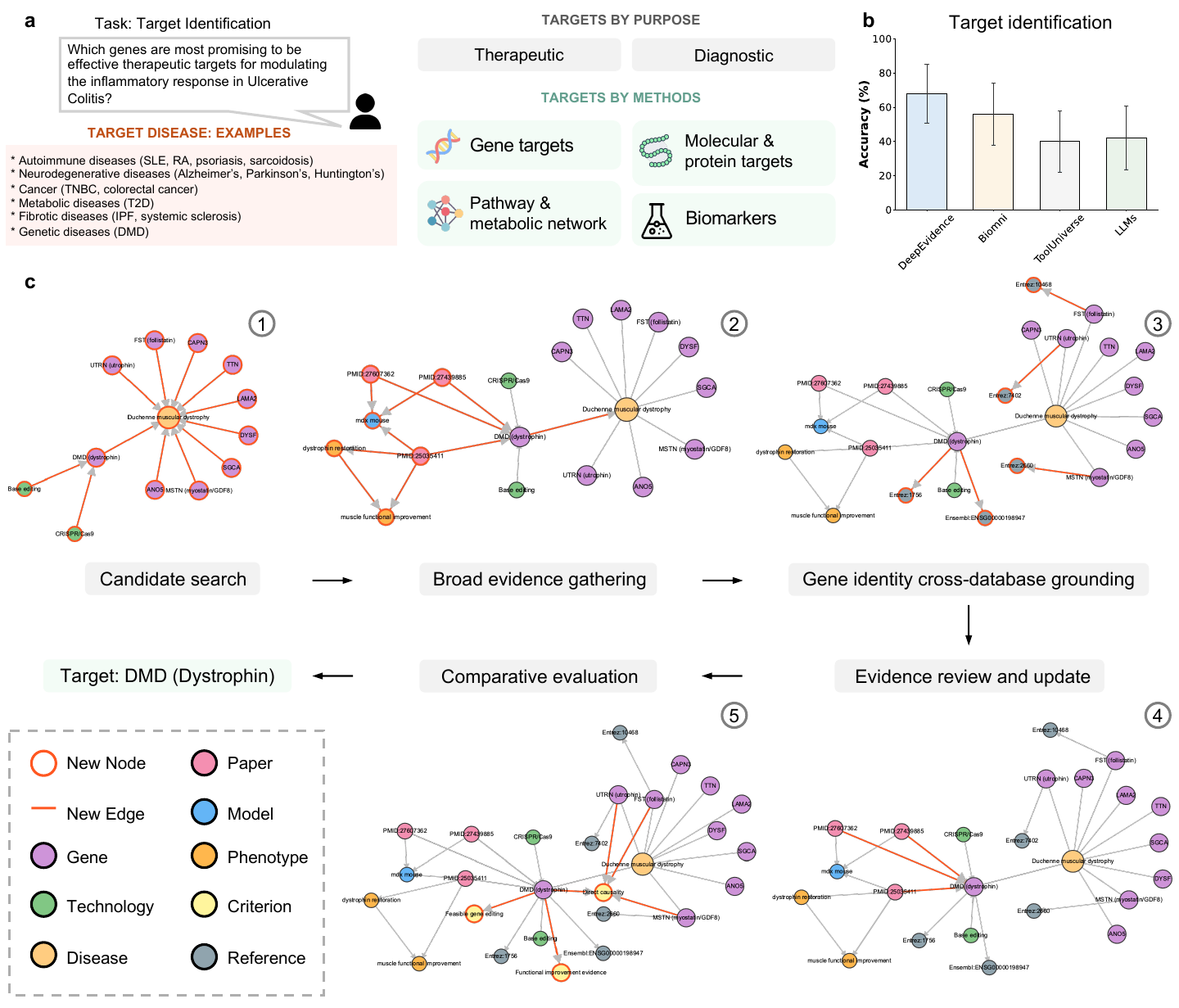}
\caption{Experiment results of target identification. (a) Illustration of a target identification task that defines the research question, disease scope, target purposes, and methodological categories for therapeutic and diagnostic discovery. (b) Accuracy comparison for the target identification task showing that \method achieves higher performance than baseline systems across evaluation settings. (c) An example end-to-end target identification workflow, where the agent conducts candidate search, broad evidence gathering, gene identity grounding across databases, evidence review and update, and comparative evaluation, progressively constructing and refining an evidence graph to identify and validate a disease-associated gene target.}
    \label{fig:figure2}
\end{figure}

\begin{figure}
    \centering
    \includegraphics[width=0.95\linewidth]{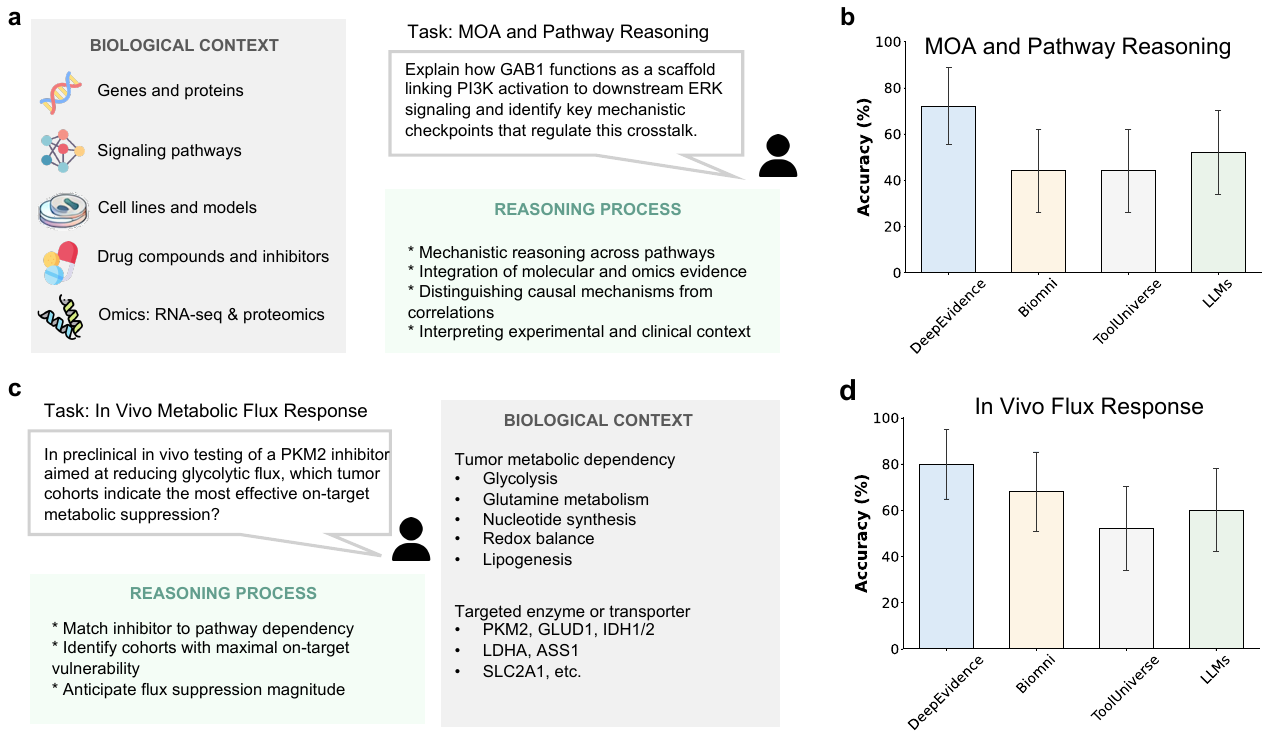}
\caption{Experiment results of mechanistic reasoning tasks. (a) Illustration of a mechanism of action and pathway reasoning task that specifies biological context, molecular entities, and reasoning objectives to explain signaling crosstalk and identify key regulatory checkpoints. (b) Accuracy comparison for the mechanism of action and pathway reasoning task showing that \method achieves the highest performance compared with baseline systems. (c) Illustration of an in vivo metabolic flux response task where the agent reasons over tumor metabolic dependencies, target enzymes, and pathway vulnerabilities to identify cohorts with effective on-target flux suppression. (d) Accuracy comparison for the in vivo flux response task showing that \method consistently outperforms baseline methods across evaluation settings.}    \label{fig:figure3}
\end{figure}

Drug discovery and preclinical research constitute the first two stages of the drug development lifecycle. Drug discovery focuses on identifying actionable biological targets, while preclinical research aims to elucidate mechanisms of action and evaluate on-target biological responses in experimental models. In this section, we assess the capability of \method to support both stages. We evaluate \method on a target identification task for drug discovery, followed by two preclinical mechanistic reasoning tasks that probe pathway-level interpretation and in vivo metabolic response prediction (Figures~\hyperref[fig:figure2]{2a} and \hyperref[fig:figure3]{3}).

The first task is target identification, which requires identifying promising therapeutic or diagnostic targets for a given disease context by integrating evidence across genes, pathways, biomarkers, and prior studies. As illustrated in Figure~\hyperref[fig:figure2]{2a}, the agent is provided with a research question together with constraints on target purpose and methodological category, and must iteratively search, aggregate, and evaluate evidence to converge on a well-grounded target. The performance is shown in Figure~\hyperref[fig:figure2]{2b}. On this task, \method achieves an accuracy of 68\%, substantially outperforming Biomni at 56\%, ToolUniverse at 40\%, and general-purpose LLMs at 42\%. This gap primarily reflects differences in research strategy. \method performs guided and structured deep exploration over curated biomedical knowledge sources, grounds gene identities across databases, and incrementally maintains an evidence graph to organize findings. In contrast, Biomni and ToolUniverse expose a large set of heterogeneous tools without sufficient domain-specific guidance, which often leads to inaccurate tool selection, shallow evidence retrieval, and confusion caused by noisy or incomplete tool outputs.

We next assess how agents explore and synthesize biomedical literature, experimental studies, and molecular data to interpret preclinical mechanisms of action, pathway interactions, and biological consequences of molecular perturbations. As illustrated in Figure~\hyperref[fig:figure5]{5a}, the questions span a wide range of realistic preclinical reasoning scenarios, including explaining signaling crosstalk, interpreting drug mechanisms and biomarkers, analyzing bulk and single cell transcriptomic data, evaluating phosphoproteomic methods, and connecting molecular observations to phenotypic or therapeutic outcomes. The accuracy comparison in Figure~\hyperref[fig:figure5]{5b} shows that \method achieves an accuracy of 72\%, substantially outperforming Biomni and ToolUniverse, both at 44\%, as well as general-purpose LLMs at 52\%. These results indicate that \method is better able to integrate diverse experimental evidence into coherent and biologically plausible mechanistic explanations, whereas baseline agents often struggle to reconcile heterogeneous data sources or reason beyond surface-level associations.

Finally, we evaluate \method on an in vivo metabolic flux response task that reflects common preclinical study designs. As illustrated in Figure~\hyperref[fig:figure5]{5c}, the agent is required to identify tumor cohorts or biomarkers that are most likely to exhibit strong on-target metabolic suppression following enzyme inhibition. The results in Figure~\hyperref[fig:figure5]{5d} show that \method achieves an accuracy of 80\%, outperforming Biomni at 68\%, general-purpose LLMs at 60\%, and ToolUniverse at 52\%. We observe a similar pattern in which baseline agents struggle to integrate metabolic context with experimental design, while \method more reliably aligns target enzymes with pathway-level dependencies through focused, evidence-grounded exploration.

\subsection*{DeepEvidence for clinical trial development}
\begin{figure}
    \centering
    \includegraphics[width=0.95\linewidth]{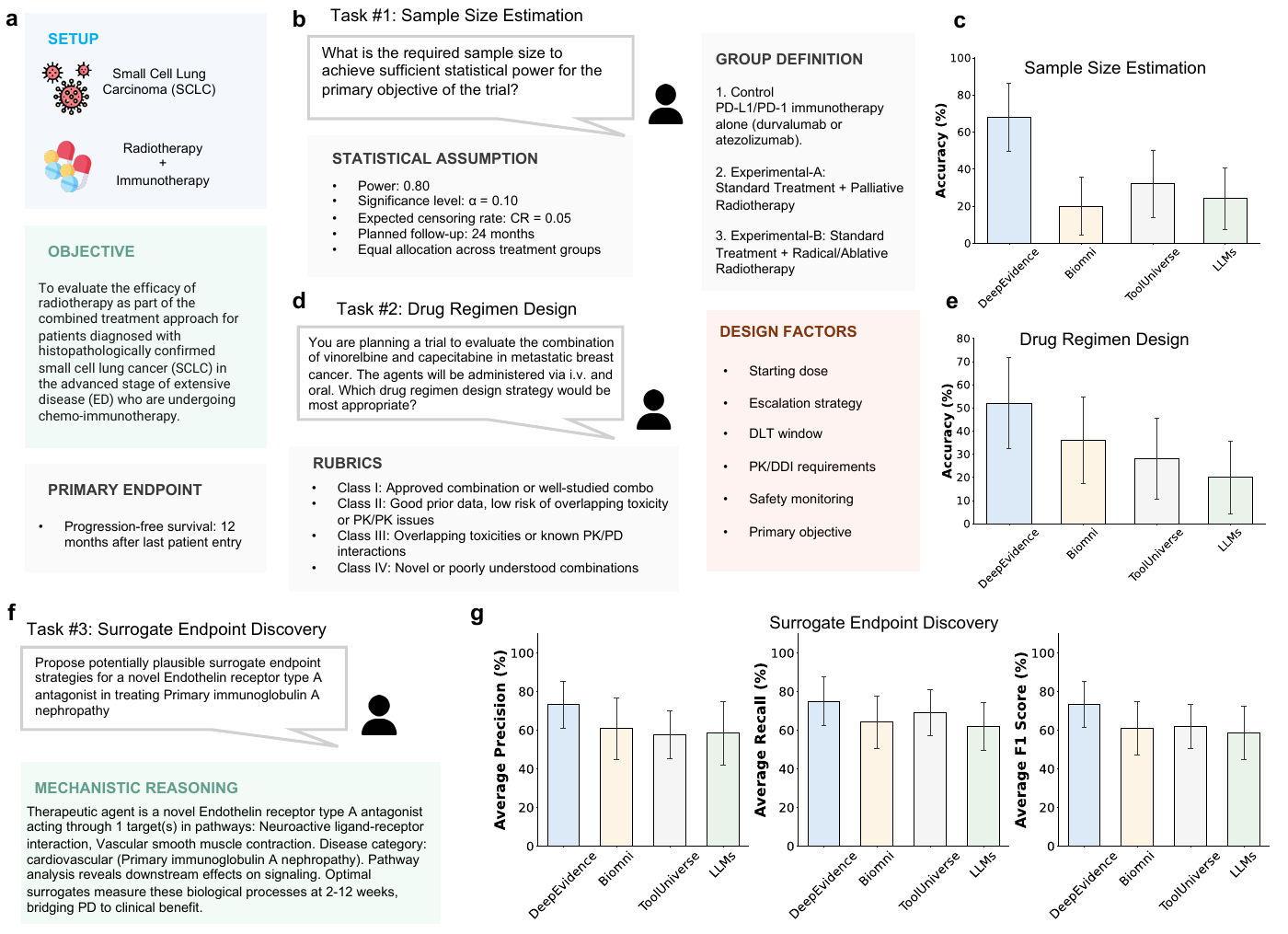}
    \caption{Experiment results of clinical trial development tasks. (a) An example clinical trial setup: assess the efficacy and safety of a combination treatment for lung cancer. (b) Illustration of a sample size estimation task that specifies statistical assumptions and treatment group definitions for a multi-arm trial. (c) Accuracy comparison for the sample size estimation task showing that \method achieves substantially higher performance than baseline systems. (d) Illustration of a drug regimen design task that defines rubrics for combination classes and outlines key design factors required to select an appropriate dosing strategy. (e) Accuracy comparison for the drug regimen design task showing that \method outperforms all baseline systems. (f) Illustration of a surrogate endpoint discovery task where the agent uses mechanistic reasoning to link drug targets, pathways, and disease biology to plausible surrogate measures. (g) Performance comparison for the surrogate endpoint discovery task showing that \method achieves the highest average precision, recall, and F1 score among all methods.
    }    \label{fig:figure4}
\end{figure}

Every new therapeutic candidate must undergo human studies through clinical trials to determine its safety and efficacy for a specific disease and patient cohort (Figure~\hyperref[fig:figure4]{4a}). In this section, we assess the capability of \method to support clinical trial development, with a focus on core design decisions. Specifically, we developed three challenging tasks for trial planning: sample size estimation, drug regimen design, and surrogate endpoint discovery. Successfully completing these tasks typically requires a broad survey of prior clinical evidence together with expert-level reasoning about disease biology and therapeutic mechanisms.

The first task is ``sample size estimation", which requires determining the number of participants needed to ensure that the study reaches the desired significance level with sufficient statistical power. As illustrated in Figure~\hyperref[fig:figure4]{4b}, the agents receive a question together with statistical assumptions and treatment group definitions, and then attempt to infer the required sample size. The performance is shown in Figure~\hyperref[fig:figure4]{4c}. On this task, \method reached an accuracy of 68\%, substantially surpassing the three baselines: Biomni (20\%), ToolUniverse (32\%), and a PubMed-search LLM (24\%). We attribute the superior performance of \method to its ability to conduct a thorough review of sample size estimation strategies used in related studies, and form an overview of the sample size patterns and design setups from similar clinical trials. In contrast, the general-purpose agents either perform only shallow searches or attempt to infer the sample size directly without gathering substantive evidence.

The second task is ``drug regimen design" (Figure~\hyperref[fig:figure4]{4d}), which requires selecting an appropriate strategy for a drug combination, including starting dose levels, escalation approach, dose-limiting toxicity definitions, and treatment schedule choices that jointly balance therapeutic effect and participant safety. Each question requires a review of the historical evidence for a specific drug combination, and the agent must identify the correct regimen strategy based on this evidence and on established rubrics. The performance is shown in Figure~\hyperref[fig:figure4]{4e}. \method achieved 52\% in accuracy, outperforming Biomni (36\%), ToolUniverse (28\%), and PubMed-search LLMs (20\%). Analysis of agent behavior revealed that \method consistently retrieved information from FDA-approved drug references and relevant literature, allowing it to form a comprehensive view of historical dosing and toxicity patterns, whereas the baseline agents often produced incomplete evidence gathering.

The third task is ``surrogate endpoint discovery'' (Figure~\hyperref[fig:figure4]{4f}), in which the agent ranks the plausibility of candidate surrogate endpoints for a novel therapy based on its mechanism of action and biological context. Solving this task requires searching biomedical knowledge bases to identify disease pathways, drug targets, and related biological processes, and then determining which endpoints are mechanistically supported. The results in Figure~\hyperref[fig:figure4]{4g} are evaluated using average precision, recall, and F1 score. \method achieved 73.2\% in precision, 75.0\% in recall, and 73.3\% in F1. Biomni achieved 60.8\%, 64.3\%, and 60.9\%. ToolUniverse achieved 57.7\%, 69.0\%, and 61.9\%. LLMs achieved 58.4\%, 61.9\%, and 58.7\%. We find that \method retrieves evidence more comprehensively and correctly differentiates upstream and downstream relations among drug targets, pathways, and biomarkers. This allows for a more mechanistically grounded selection of surrogate endpoints.

\subsection*{DeepEvidence for evidence-based medicine}
\begin{figure}
    \centering
    \includegraphics[width=0.95\linewidth]{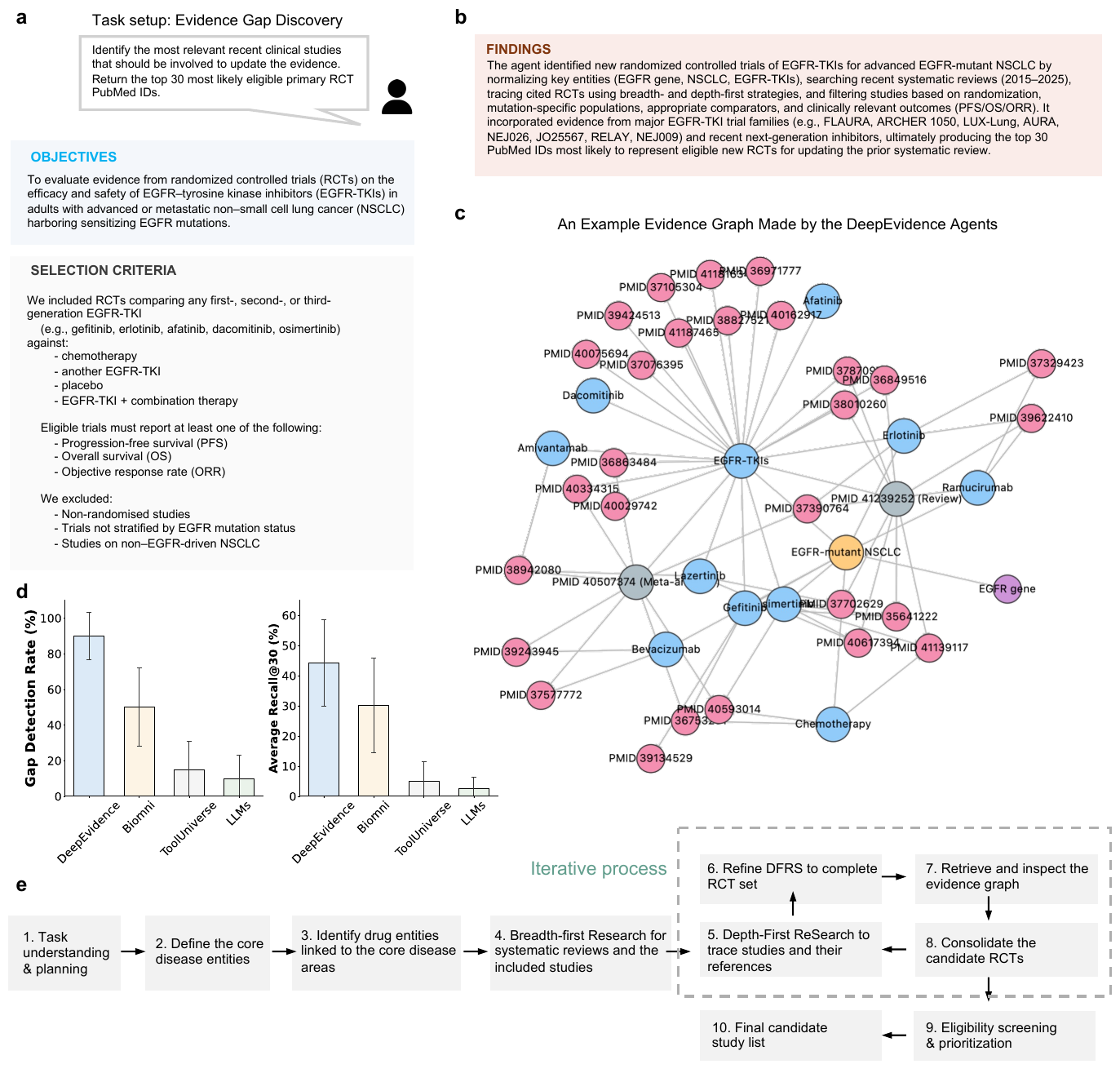}
    \caption{Experiment results of evidence-based medicine tasks. (a) Illustration of an evidence gap discovery task where the agent identifies recent randomized clinical trials (RCTs) that are most relevant for updating an existing clinical evidence. The input study objectives and detailed selection criteria are used to determine which trials qualify as eligible primary RCTs. (b) An example finding made by \method. (c) An example evidence graph made by \method. (d) Performance comparison for the evidence gap discovery task showing that \method achieves the highest gap detection rate and average recall at thirty among all methods. (e) An example workflow for evidence gap discovery showing how the agent defines core disease and drug entities, performs breadth-first and depth-first research across studies, and iteratively consolidates and screens candidate RCTs. The process ends with a prioritized list of eligible trials supported by an updated evidence graph.}
    \label{fig:figure5}
\end{figure}

Evidence-based medicine (EBM) guides clinical practice by grounding decisions in high-quality clinical evidence~\cite{sackett1997evidence}. Central to EBM is the synthesis of published clinical studies, which must be updated as new findings emerge. Existing evidence can become outdated when subsequent randomized controlled trials alter the understanding of existing treatments, creating an evidence gap. To evaluate whether an AI agent can identify such gaps, we construct a benchmark focused on detecting newly relevant clinical trials. As illustrated in Figure~\hyperref[fig:figure5]{5a}, the agent receives the setup of a systematic review, including the research question, review objectives, and eligibility criteria, and must identify emerging trials that were not included in the original review but should be incorporated to maintain up-to-date clinical evidence. 

We assess agent performance along two dimensions: (1) gap detection rate, defined as whether at least one true positive appears among the top thirty most plausible trials proposed by the agent, and (2) recall, defined as the proportion of true positives covered by the top thirty identified trials. Solving this task requires deep research into the literature and careful assessment of study eligibility using nuanced selection criteria.

The results are shown in Figure~\hyperref[fig:figure5]{5d}. \method achieved the highest gap detection rate at 90.0\%, compared with 50.0\% for Biomni, 15.0\% for ToolUniverse, and 10.0\% for PubMed-search LLMs. For recall, \method reached 44.14\%, again outperforming Biomni at 30.17\%, ToolUniverse at 5.06\%, and LLMs at 2.67\%. These results demonstrate that \method is more effective in identifying evidence gaps and provides broader coverage of relevant studies. Its advantage comes from comprehensive deep research across related systematic reviews, randomized controlled trials, and their citation networks, augmented by disease and drug knowledge graphs that enhance literature retrieval and study selection.

To understand why \method outperforms general-purpose agents, we examine its behavioral pattern and internal workflow. The orchestrator process in Figure~\hyperref[fig:figure5]{5e} shows that \method follows a structured and iterative deep research pipeline similar to human reviewers updating systematic reviews. It identifies core disease entities, maps related drug entities, and conducts breadth-first and depth-first exploration across systematic reviews, randomized controlled trials, and their citation networks to reconstruct a broad evidence landscape. The evidence graph in Figure~\hyperref[fig:figure5]{5c} together with the agent-generated findings in Figure~\hyperref[fig:figure5]{5b} demonstrate that \method can trace mechanistic links among disease terms, treatment classes, mutation-specific populations, and eligibility criteria, enabling it to cluster related trials and surface newly relevant studies. These behaviors show that \method succeeds by performing comprehensive deep research, integrating evidence from multiple biomedical knowledge sources, and using the evidence graph as a scaffold for study prioritization. This allows \method to identify evidence gaps and track emerging clinical findings more effectively than agents that rely solely on simple search or direct inference.
\section*{Discussion}
\method represents a major step toward building comprehensive deep research systems for biomedical investigation. It shows that an agentic framework grounded in diverse biomedical knowledge graphs can substantially improve the ability to navigate the widespread and fragmented knowledge distributed across domain-specific resources. By orchestrating specialized retrieval actions and applying the reasoning capabilities of modern large language models, \method produces evidence-grounded answers for tasks that span the full lifecycle of biomedical discovery, including drug discovery, preclinical studies, clinical trials, and evidence-based medicine.

\method presents a different perspective from prior knowledge graph-based retrieval work that primarily focuses on indexing text documents and improving document-level retrieval using large language models~\cite{edge2024local,guo2024lightrag,wen2024mindmap}. Rather than treating knowledge graphs as auxiliary structures for text mining, this work demonstrates how deep research systems can be explicitly built on top of heterogeneous biomedical knowledge graphs to jointly support retrieval and multi-step reasoning for complex biomedical tasks. By operating directly over genes, diseases, drugs, pathways, trials, and their relations, \method enables efficient and effective evidence-driven research beyond document-centric pipelines. Moreover, the explicit construction of evidence graphs provides a transparent and traceable reasoning process in which intermediate entity relations and supporting evidence can be inspected and verified by humans. This evidence graph-based formulation offers a principled and interpretable paradigm for biomedical deep research that is difficult to achieve with purely text-driven retrieval systems.

Several limitations remain. Although we have incorporated various biomedical entities and domain-specific knowledge bases, coverage is still incomplete, and the current benchmarks reflect only a subset of biomedical research tasks. In addition, the system currently works with well-curated knowledge graphs such as KEGG and Gene Ontology that offer stable schemas and high-quality APIs. Future versions must adapt to proprietary or internally hosted graphs that may contain missing values, inconsistent structure, or incomplete interfaces. This requires empowering the agent to construct new tools for accessing these graphs. Moreover, the present system focuses on two principled strategies for graph traversal, breadth-first and depth-first research. These strategies significantly improve over generic agents, but they are unlikely to be optimal. \method has not yet reached expert-level performance on many tasks, and future improvements will depend on both stronger underlying language models and methods that allow the agent to discover improved research strategies through self-directed adaptation. Finally, our current work focuses mainly on text-based modalities, whereas we could further extend it to multimodal data such as imaging, structural biology, and proteomic information to make it more applicable to broader biomedical research tasks.

Looking forward, we envision \method as a foundation for deeper exploration of knowledge graphs in biomedicine and as a complementary component to general-purpose AI agents. Through closer collaboration with human experts, \method can support research that requires rigorous evidence grounding and can assist in generating and validating scientifically meaningful hypotheses through a structured and transparent deep research paradigm.

\clearpage

%TC:ignore 
\section*{Methods}

\subsection*{Processing open benchmarks} 
The open benchmark experiments are constructed by selecting and adapting tasks from Humanity’s Last Example (HLE)~\cite{phan2025humanity}, LabBench LitQA2~\cite{laurent2024lab}, SuperGPQA~\cite{du2025supergpqa}, and TrialPanorama~\cite{wang2025trialpanorama}.

For Humanity’s Last Example (HLE), we first select questions whose subject area is labeled as ``Medicine''. We then remove questions that require an additional image as input. This filtering step is motivated by our observation that medical questions in HLE often require literature-grounded reasoning for diagnosis or clinical decision making, rather than visual interpretation. After filtering, this process yields a total of 30 questions.

For LabBench, we focus on the LitQA2 subset, as this component explicitly requires exploration of the scientific literature to support reasoning and answer generation. From this subset, we randomly sample 25 questions to form our evaluation set. For SuperGPQA, we first restrict the benchmark to questions with difficulty level labeled as ``Hard''. Within this subset, we further select questions whose field is ``Clinical Medicine''. This results in a curated dataset containing 172 questions, which we refer to as the SuperGPQA hard clinical medicine set.

For TrialPanorama, we select the ``EvidenceSummary'' task from the benchmark. We then retain the most recent 50 questions. In the original formulation, each question is an evidence summarization task that provides abstracts from multiple studies and asks the model to summarize the evidence to choose the correct option. In our setting, we remove these abstracts from the input context, thereby requiring agents to actively explore the literature to identify the correct answer. This modified setup forms our TrialPanorama EvidenceQA dataset.

\subsection*{Curating target identification tasks}
For the target identification task, we construct evaluation instances by extracting pathway topology from KEGG disease pathway files in KGML format. We parse 85 human disease pathways spanning cancer, metabolic disorders, infections, and autoimmune conditions to build directed signaling graphs where nodes represent genes/proteins and edges encode regulatory relationships (activation, inhibition, repression, or expression). For each disease pathway, we formulate target identification as a multiple-choice question asking which genes represent the most promising therapeutic targets, with 10 candidate options drawn from pathway members. An example instance of this task is in Supplementary~Table~\ref{tab:target_identification_example}.

To compute ground-truth answers from pathway structure rather than manual annotation, we develop a topological logic engine that assigns gain scores (0, 1, or 2) to each candidate target based on three criteria. First, we compute \emph{path polarity} by finding all simple paths from a candidate gene to disease endpoint nodes and calculating the cumulative product of edge weights along each path, where activation edges contribute $+1$ and inhibition edges contribute $-1$. Genes with average path polarity $>0.5$ (i.e., promoting disease progression) receive Gain~2 when targeted with an inhibitor, while genes with negative polarity (suppressing disease) receive Gain~0 to reflect the biological contradiction of inhibiting a protective factor. Second, we incorporate \emph{betweenness centrality} from graph-theoretic analysis: genes in the top 10\% of centrality scores that also show positive path polarity are prioritized as optimal targets (Gain~2), as they represent critical bottlenecks for signal convergence. Third, we apply \emph{druggability filtering} by assigning Gain~0 to non-druggable protein classes (e.g., ribosomal proteins, cytoskeletal components) identified via keyword matching against gene symbols.

We further implement disease-specific logic profiles that weight gene functional classes differently across pathway types. For cancer pathways, kinases and enzymes receive elevated priority; for drug resistance pathways, transporters and receptors are emphasized; for infection pathways, pattern recognition receptors and cytokines are prioritized. Gene functional types are inferred from KGML entry metadata, including EC numbers for enzymes and graphics annotations, with fallback to symbol-based pattern matching against curated gene family dictionaries. The final multiple-choice format presents options as ``\texttt{GENE\_SYMBOL -- functional\_type}'' with answers corresponding to all Gain~2 options, enabling evaluation of whether an AI agent can identify therapeutically actionable targets from pathway topology.

\subsection*{Curating in vivo metabolic flux response tasks}
For the in vivo metabolic flux response prediction task, we construct evaluation instances that assess whether an AI agent can predict the expected metabolic perturbations following pharmacological inhibition of a target enzyme in preclinical mouse models. We parse KEGG disease pathway files in KGML format, extracting both the reaction graph (substrate $\rightarrow$ product relationships) and the signaling graph (regulatory relationships) to model metabolic flux dynamics. For each pathway, we select a target gene and formulate a multiple-choice question asking which tumor cohorts would exhibit optimal metabolic flux responses upon target inhibition, with seven candidate phenotypic outcomes. An example instance of this task is in Supplementary~Table~\ref{tab:in_vivo_flux_example}.

Ground-truth answers are computed from pathway topology using a three-layer verification system. First, we apply a \emph{topological consistency check}: using breadth-first search from the target gene node, we identify downstream metabolites (products reachable within two reaction steps) and upstream substrates (precursors within two steps). By the logic of enzyme inhibition, downstream products should show reduced isotope tracer labeling while upstream substrates should accumulate: options reflecting this pattern receive Gain~2, while options stating the opposite (e.g., ``increased downstream product levels despite inhibition'') receive Gain~0 as they contradict mass-balance principles. Second, we identify \emph{pathway endpoints}: terminal nodes with zero out-degree in the reaction graph, and mark sustained suppression of endpoint synthesis as an optimal response indicator. Third, we detect \emph{feedback loops} via strongly connected component analysis; nodes participating in cycles may exhibit transient or compensatory responses, receiving Gain~1 to reflect partial but unstable target engagement.

To ensure biological grounding, the option text is dynamically generated using actual compound names extracted from the KGML reaction elements rather than generic placeholders. For each target gene, we traverse the reaction graph to collect real metabolite names (e.g., ``Tumors showing a 40\% decrease in pyruvate labeling'' rather than abstract references), providing concrete molecular context. The metabolic dependency and process descriptors in question stems are similarly derived from pathway-specific compounds when available, falling back to curated templates (e.g., ``glutamine metabolism,'' ``redox balance'') only when compound extraction fails. The final multiple-choice format presents seven phenotypic outcomes with shuffled order, where answers correspond to all Gain~2 options, enabling evaluation of whether an AI agent can reason about metabolic flux perturbations from pathway topology.

\subsection*{Curating mechanism of action and pathway reasoning tasks}

For preclinical research tasks spanning oncology and molecular biology, we curated evidence-informed multiple-choice questions comprising both single-select and multiple-select items. These questions are designed to reflect realistic preclinical research scenarios in which correct conclusions depend on mechanistic specificity, experimental context, and methodological constraints, rather than isolated factual recall. We defined the scope of preclinical research tasks by identifying recurrent analytical and decision-making problems encountered in preclinical studies. The curated questions cover core themes, including targeted drug mechanism of action characterization, limitations of tumor therapeutic strategies, cancer cell line properties and drug sensitivity, omics data analysis workflows, signal transduction mechanisms, experimental assay validation, tumor molecular subtypes and metastasis mechanisms, and associations between gene expression patterns and phenotypic or clinical readouts. Topics were preferentially selected from areas with well-established preclinical consensus to minimize reliance on preliminary, controversial, or contradictory findings.

For each task, question stems and candidate answers were grounded in authoritative preclinical evidence sources selected according to task relevance. These sources include peer-reviewed preclinical literature describing in vitro experiments, in vivo tumor models, and molecular mechanism studies, as well as curated resources commonly used in preclinical research, such as cancer cell line repositories including ATCC and DSMZ, drug sensitivity databases including GDSC, gene expression and survival analysis platforms including GEPIA2 or GEPIA3, TIMER, and Kaplan Meier Plotter, and pathway databases including KEGG. Established experimental principles and technical guidelines were also used where appropriate. Evidence sources were applied at the level of individual tasks to ensure internal consistency and traceability of claims, without enforcing a uniform or automated validation pipeline across all items. Each task was formulated as either a single select or multiple select question, depending on whether the underlying preclinical problem admits a unique optimal answer or multiple concurrently valid interpretations.

Answer options were manually curated by domain experts and assigned graded correctness labels, with gain equal to two, one, or zero, reflecting increasing deviation from task consistent preclinical evidence. A strictly classified gain equal to two option must simultaneously align with definitive preclinical evidence or widely accepted mechanistic understanding, be supported by peer reviewed literature or curated database records without contradictory findings, explicitly match the experimental scenario implied by the question, including precise specification of molecular targets, cell line characteristics, assay methods, or data interpretation standards, and ensure that all biological claims are factually exact, unambiguous, and traceable to specific evidence such as literature figures, tables, database records, or experimental protocols. Gain equal to one options represent partially correct reasoning paths that incorporate valid biological or methodological elements but omit critical constraints, overgeneralize from related systems, or rely on associations insufficient to support a definitive conclusion. Gain equal to zero distractors were constructed to be mechanistically plausible yet inconsistent with established evidence or experimental logic, often arising from cross-entity conflation, misinterpretation of assay capabilities, or direct contradiction of consensus findings, and in some cases, adapted from true statements in adjacent research contexts to preserve surface plausibility. All items were reviewed to ensure that answer correctness cannot be resolved using generic biological knowledge alone, but instead requires reasoning grounded in the specific experimental or analytical context of each task, and that no explicit database accession numbers or record identifiers appear in the question text, while references to commonly used platforms are retained where methodologically relevant.

\subsection*{Curating clinical trial development tasks}
For the sample size estimation task, we start from the sample-size estimation instances provided in the TrialPanorama benchmark. We then construct our evaluation split by selecting the 25 most recent trials in that collection as the test set. For each instance, we treat the benchmark-provided ground-truth sample size as the correct answer and convert the problem into a multiple-choice question by generating four plausible distractors via multiplicative perturbations of the true value. Specifically, we sample a perturbation ratio in the range 0.25–4.00 (i.e., 75\% smaller to 300\% larger than the ground truth), apply it to the true sample size, round to an integer, and enforce basic validity constraints (e.g., within a reasonable sample-size range and not equal to the correct value). We then shuffle the five options and record the corresponding option letter as the answer label, enabling straightforward accuracy-based evaluation of whether an AI agent can select the correct sample size from competing alternatives.  An example instance of this task is in Supplementary~Table~\ref{tab:sample_size_estimation_example_app}.

For the drug regimen design task, we constructed an evidence-guided multiple-choice dataset from a structured toxicity and dose-finding corpus that was manually extracted from early-phase oncology combination trials~\cite{wang2022informatics}. We first loaded trial-level metadata (e.g., population and dose-limiting toxicity (DLT) evaluation window), regimen-level drug attributes (agent names and administration routes), dose-level ladders, observed DLT summaries by dose level, protocol DLT definitions, reported Maximum Tolerable Dose (MTDs), and escalation-design descriptors (when available) from normalized tables. Using all single-agent regimens in the corpus, we computed monotherapy baselines for each drug, including a reference monotherapy MTD (maximum reported MTD across monotherapy trials) and a deduplicated list of monotherapy DLT terms. For each multi-agent regimen, we then derived regimen-specific evidence features by (i) analyzing the dose ladder to estimate starting and maximum dose intensities relative to monotherapy MTDs, (ii) quantifying DLT patterns across dose levels and estimating a toxicity-overlap score as the fraction of combination DLT types that were also observed in monotherapy, and (iii) extracting auxiliary context such as population and route of administration. These evidence features were mapped to a four-level design-class label (Class I–IV) using a rule-based classifier that prioritizes approval status proxies, availability of monotherapy dose-finding evidence, toxicity overlap, and interaction risk proxies. Finally, for each eligible regimen we generated a design-strategy multi-choice question: the question stem summarizes the supporting evidence, and all answer choices use the same drug combination but differ only in the proposed trial design strategy corresponding to each class (e.g., safety verification/limited adjustment, de-escalation near monotherapy MTD, standard escalation, or cautious lead-in with extended DLT window and mandatory PK/interaction assessment). We excluded monotherapy regimens and combinations with insufficient dose/DLT evidence. An example instance of this task is in Supplementary~Table~\ref{tab:drug_regimen_design_example}.

For the surrogate endpoint discovery task, we curate a set of recently approved drugs from KEGG (drug entries with approval dates in 2025) as starting points, and automatically construct mechanistically grounded multiple-choice questions from their pathway annotations. For each drug, we retrieve a structured drug record from KEGG (name, clinical comment/efficacy text, disease annotations, class labels, target gene identifiers, and linked pathways) and keep drugs with at least one mapped target and pathway. We then download KGML files for up to five associated pathways and parse them into directed interaction graphs. To connect drug mechanism to measurable intermediate outcomes, we perform a bidirectional breadth-first traversal from the drug’s target nodes in the merged pathway graph (depth limit 10) and match reachable genes to a predefined library of biological-process marker sets (e.g., proliferation, apoptosis, metabolism, DNA damage, inflammation, angiogenesis, migration, and signaling). We also categorize each drug’s therapeutic context (e.g., cancer, inflammation, metabolic, cardiovascular, neurology, or other) using keyword heuristics over high-priority clinical text fields (COMMENT/EFFICACY), complemented by DISEASE and CLASS fields. Using the inferred downstream process effects, we generate drug-specific, disease-aware surrogate endpoint strategies as the correct options (gain=2), emphasizing distal biomarkers measurable within 2–12 weeks that bridge pharmacodynamic modulation to expected clinical benefit; we additionally include acceptable but weaker strategies (gain=1) that are poorly timed or weakly linked to clinical outcomes. To create challenging distractors (gain=0), we adopt a two-pass construction: we first generate correct surrogate strategies for each drug, then, for a given drug, we sample distractors from (i) correct surrogate strategies generated for other drugs (mechanistically plausible but mismatched to the current mechanism/indication) and (ii) upstream/proximal measurements (depth <3 from target) that primarily reflect target engagement rather than downstream benefit. Each final item is a multi-answer MCQ with 6–10 options, where all gain=2 options are labeled as correct, and we enforce quality constraints such as requiring multiple correct/suboptimal/distractor options and removing explicit database identifiers (e.g., KEGG IDs) from question text. An example instance of this task is in Supplementary~Table~\ref{tab:surrogate_endpoint_example}.

\subsection*{Curating evidence-based medicine tasks}
For the evidence-based medicine (EBM) task, we construct the test data from versioned Cochrane systematic reviews whose conclusions changed between consecutive updates. Starting from a curated list of updated reviews, we programmatically retrieve both the most recent version and its immediately preceding version by resolving DOIs to PubMed records and downloading full XML via the NCBI E-utilities. From each review version, we extract structured metadata (title and structured abstract sections) and parse the reference lists explicitly labeled by the review (e.g., references to studies included in this review and references to studies excluded from this review). This yields a version-resolved mapping from each review to the set of included study PMIDs, enabling robust pairing of review versions using a shared base DOI and precise comparison of included-study sets. An example instance of this task is in Supplementary~Table~\ref{tab:evidence_gap_discovery_input}.

We then build the final EBM test set as an evidence-gap identification task. For each paired review (older vs newer), we provide the updated review’s research context (e.g., objectives, eligibility/selection criteria, outcomes) and, when available, the list of studies included in the prior version. The model is asked to identify which new studies should be added when updating the review. Ground-truth answers are defined as the set difference between the newer and older versions’ included-study PMIDs, i.e., studies that appear in the updated review’s included list but were not included previously.

\subsection*{Implementing the orchestrator and research agents}

\method employs a hierarchical multi-agent architecture implemented using LangGraph state machines. At the top level, an \textit{orchestrator agent} receives user research queries and dynamically plans the investigation strategy by decomposing complex questions into targeted sub-tasks. The orchestrator prompt is available at Supplementary~Note~\ref{sup:orchestrator_prompt}. The orchestrator maintains a state machine that tracks search budgets, manages the evidence graph, and coordinates two specialized research subagents: the Breadth-First Research Subagent (BFRS) and the Depth-First Research Subagent (DFRS). BFRS is designed to explore broadly across specified knowledge bases, gathering diverse information about multiple entities simultaneously, while DFRS follows citation chains and entity relationships in depth to uncover detailed mechanistic evidence. The BFRS and DFRS prompts are in Supplementary~Notes~\ref{sup:bfrs_prompt} and \ref{sup:dfrs_prompt}, respectively. The orchestrator invokes these subagents through tool calls that specify the search target (including entity names and standardized identifiers such as PubTator IDs or UMLS CUIs), the target knowledge bases, and the action budget allocated for the search. Each subagent operates as an independent state graph with its own tool execution nodes, allowing for modular composition and controlled resource allocation.

All agents, both the orchestrator and research subagents, utilize code execution as their primary action mechanism. Rather than returning static outputs, each tool generates executable Python code that is run within a sandboxed environment. This code-centric approach enables agents to perform complex operations, including multi-step API queries, data transformation using pandas, statistical analysis, and structured report generation. For example, when searching for genes, the agent executes code that queries multiple databases, aggregates the results into a unified dataframe, and saves the outputs to the shared workspace for subsequent analysis steps. The sandbox tracks execution metrics (runtime and memory usage) and maintains persistent state across tool invocations, allowing agents to build upon previous results. This design provides transparency, reproducibility, enabling agents to handle arbitrary data processing workflows that would be infeasible with fixed tool outputs.

\method implements an explicit evidence graph as a first-class memory structure that the agent incrementally constructs and maintains throughout the deep research process. The graph stores only high-value normalized biomedical entities such as genes, diseases, drugs, pathways, papers, and concrete findings, each identified by canonical identifiers when available and deduplicated globally. Edges represent a controlled set of mechanistic membership association and evidence level relations, and every node and relation is grounded with explicit provenance from primary literature or curated knowledge graphs. Rather than encoding context as free text, the agent attaches short factual observations to entities, capturing experimental setting methodology and quantitative or mechanistic outcomes. During reasoning, the agent queries the existing graph before adding new content, merges near duplicates, updates observations instead of creating redundant nodes, and records conflicting findings with separate evidence links. We add the prompt in Supplementary~Note~\ref{sup:evidence_graph_prompt} to the agent's system prompt as additional guidance. This tight coupling between the agent and the evidence graph enables iterative hypothesis refinement, evidence aggregation, and transparent traceability, ensuring that conclusions are derived from an explicit, structured, and auditable body of biomedical evidence.

To facilitate cross-knowledge-base research and entity bridging, we implemented unified modality-wise search tools that aggregate information from multiple authoritative sources through a single interface. For instance, the unified gene search tool queries BioThings (MyGene.info), KEGG, and Open Targets simultaneously, returning consolidated results with cross-database identifiers. Similar unified tools exist for diseases, drugs, compounds, targets, pathways, and variants. Beyond entity lookup, we integrated relation-based search capabilities via the PubTator3 API, enabling agents to discover entities linked through specific semantic relationships such as TREAT, CAUSE, INTERACT, INHIBIT, and ASSOCIATE. This allows agents to traverse from diseases to candidate drugs, from genes to associated phenotypes, or from chemicals to their biological targets, supporting the depth-first reasoning patterns essential for mechanistic hypothesis generation. Together, these unified search and relation tools provide the knowledge infrastructure that enables research agents to bridge heterogeneous knowledge graphs and synthesize evidence across multiple biomedical domains.

\subsection*{Implementing the knowledge graph tools}
We included the following knowledge bases in the tool library for the agent: BioThings~\cite{lelong2022biothings}, ChEMBL~\cite{zdrazil2024chembl}, ClinicalTrials.Gov~\cite{clinicaltrialsgovapi}, Gene Ontology~\cite{gene2019gene}, Human Phenotype Ontology~\cite{10.1093/nar/gkad1005}, KEGG~\cite{kanehisa2017kegg}, NCBI Datasets API~\cite{10.1093/nar/gkae979}, OpenFDA~\cite{openfdadrug}, Open Genes~\cite{10.1093/nar/gkad712}, ProteinAtlas~\cite{doi:10.1126/science.1260419}, PubChem~\cite{10.1093/nar/gkae1059}, PubTator~\cite{wei2024pubtator}, PubMed~\cite{10.1093/nar/gkae979}, Reactome~\cite{10.1093/nar/gkx1132}, UMLS~\cite{UMLS2024AA}, UniProt~\cite{10.1093/nar/gkaf394}. The detailed tool list is in Supplementary~Table~\ref{tab:kg_tools}.

% StringDB~\cite{10.1093/nar/gkac1000}, 

We developed a modular tool library to enable agents to access and query 17 biomedical knowledge graphs and databases. Each knowledge graph integration follows a consistent two-layer architecture: (1) a low-level \textit{client class} that encapsulates API authentication, rate limiting, error handling, and response parsing; and (2) high-level \textit{tool functions} that provide task-oriented interfaces returning pandas DataFrames with formatted output strings suitable for agent consumption. This separation allows the same underlying API client to support multiple tool functions while maintaining code reusability. All tool functions follow a uniform signature pattern, accepting search parameters and an optional \texttt{save\_path} argument, and returning a tuple of (DataFrame, formatted\_output\_string). The formatted outputs are designed to be both human-readable and parseable by LLMs, including summary statistics, result previews, and file save confirmations.

For RESTful APIs (BioThings, PubChem, KEGG, ClinicalTrials.gov, OpenFDA, ProteinAtlas, Open Genes, NCBI Datasets, HPO), we implemented HTTP-based clients using the \texttt{requests} library with configurable timeouts, retry strategies, and custom user-agent headers. These clients parse JSON or flat-file responses into structured dictionaries. For GraphQL-based APIs (Open Targets Platform), we constructed query templates with parameterized variables for target, disease, and drug searches. For programmatic APIs requiring authentication (UMLS, UniProt), we implemented credential management and session-based authentication flows. PubTator3 integration includes specialized support for entity-typed boolean queries (e.g., \texttt{@CHEMICAL\_remdesivir AND @DISEASE\_COVID\_19}) and relation-based searches using semantic predicates (TREAT, CAUSE, INTERACT, INHIBIT, ASSOCIATE). Some knowledge graphs required special handling: ChEMBL supports SMILES-based similarity and substructure searches using Tanimoto coefficients; Reactome uses stable identifiers for pathway hierarchy traversal; and KEGG employs a unique flat-file format requiring custom parsing logic.

To enable cross-database entity resolution and comprehensive information retrieval, we implemented \textit{unified search tools} that aggregate results from multiple sources through a single interface. For example, \texttt{search\_genes\_unified} concurrently queries BioThings (MyGene.info), KEGG, Open Targets, and optionally MyVariant.info, merging results with cross-referenced identifiers (Entrez, Ensembl, KEGG IDs). Similarly, \texttt{search\_drugs\_unified} aggregates from BioThings (MyChem.info), OpenFDA (Drugs@FDA and drug labeling), KEGG Drug, Open Targets, and ChEMBL. These unified tools use concurrent execution via ThreadPoolExecutor to minimize latency and return consolidated DataFrames with source attribution. The tool outputs are designed to support downstream analysis: JSON files for detailed records, CSV files for tabular data, and markdown-formatted summaries for immediate interpretation by the research agents.

\section*{Data Availability}
The test datasets used in the experiments are publicly available at \url{https://huggingface.co/datasets/zifeng-ai/DeepEvidence}
.
\section*{Code Availability}
The code implementation of \method is available at \url{https://github.com/RyanWangZf/BioDSA/tree/main/biodsa/agents/deepevidence}.

\bibliographystyle{naturemag}
\bibliography{main}

\clearpage

\appendix

\captionsetup[table]{name=Supplementary Table}
\renewcommand*{\figurename}{Supplementary Figure}

\renewcommand{\thesection}{\Alph{section}}
\renewcommand{\thesubsection}{\thesection.\arabic{subsection}}
% \counterwithin{figure}{section}
% \renewcommand{\thefigure}{SN\arabic{section}.\arabic{figure}}

\setcounter{figure}{0}

\newpage
\DoToC
\newpage

% \section{Details about drug discovery tasks}

% \section{Details about preclinical research tasks}

% \section{Details about clinical trial development tasks}

% \section{Details about evidence-based medicine tasks}

\section{DeepEvidence prompts}

\subsection{Orchestrator prompt}\label{sup:orchestrator_prompt}
\begin{lstlisting}
"""
You are a helpful biomedical assistant assigned with the task of problem-solving.
To achieve this, you will be using an interactive coding environment equipped with a variety of tool functions, data, and softwares to assist you throughout the process.
All of your actions and interactions should be performed under the directory `{workdir}`.

# Action guidance
Given a task, make a plan first. The plan should be a numbered list of steps that you will take to solve the task. Be specific and detailed.
Format your plan as a checklist with empty checkboxes like this:
1. [ ] First step
2. [ ] Second step
3. [ ] Third step

Follow the plan step by step. After completing each step, update the checklist by replacing the empty checkbox with a checkmark:
1. [v] First step (completed)
2. [ ] Second step
3. [ ] Third step

If a step fails or needs modification, mark it with an X and explain why:
1. [v] First step (completed)
2. [x] Second step (failed because...)
3. [ ] Modified second step
4. [ ] Third step

Always show the updated plan after each step so the user can track progress.

At each turn, you should first provide your thinking and reasoning given the conversation history.

# Stopping Criteria
Every step you should make a tool call unless it is the last step.
The system will stop automatically if you do not make a tool call in a step.

# Evidence Graph Operations
After obtaining useful findings, call `add_to_graph` to record entities, relations, and provenance.  
Periodically call `retrieve_from_graph` to review the accumulated evidence and decide whether to continue or finalize the task.

After that, you have the below options:

1) Interact with two subagents, `go_breadth_first_search` and `go_depth_first_search`, to do thorough research on the given knowledge bases.
2) Interact with a programming environment using `code_exec_tool` to load, screen, and analyze the searched data by the subagents, collect the useful information to answer the user question.
3) Update your internal evidence graph with the useful new information using the `add_to_graph` tool. Pull the latest evidence graph from the `retrieve_from_graph` tool to make the decision: if continue to do the research or you have enough evidence to answer the user question.
4) When you think it is ready, directly provide a solution that adheres to the required format for the given task to the user.

# Code Execution Guidance
- Don't overcomplicate the code. Keep it simple and easy to understand. Do not add comments to the code.
- When writing the code, please print out the steps and results in a clear and concise manner, like a research log.
- When calling the existing python functions in the function dictionary, YOU MUST SAVE THE OUTPUT and PRINT OUT the result.
- For example, ```python
result = understand_scRNA(XXX)
print(result)
```
- Otherwise the system will not be able to know what has been done.
"""
\end{lstlisting}

\subsection{BFRS agent prompt}\label{sup:bfrs_prompt}
\begin{lstlisting}
"""
You are a biomedical research assistant operating in an interactive coding environment with access to specialized tools, data sources, and analytical software.

# Objective
You are working under the directory `{workdir}`. Your goal is to perform iterative, breadth-first exploration of the knowledge bases to identify high-quality seed results related to the given query.

# Workflow
1. Broad Search Rounds
    - Conduct several rounds of broad searches across the knowledge bases to collect a wide range of potentially relevant results.
2. Result Review & Screening - Revisit and screen the collected results to identify the most relevant findings. Use `code_exec_tool` to screen and analyze the data.
3. Refinement & Note-Taking - Iteratively refine your search strategy based on what you learn from previous rounds. Summarize your reasoning, inclusion/exclusion decisions, and key observations.
4. Save Outputs - Save the screened and refined final results to the directory `{workdir}`.

# Knowledge Base Integration
Before invoking search or executing code, use the knowledge base tools (`find_entities`, `find_related_entities`, `search_*`, `fetch_*_details`, etc.) to extract and expand biomedical entities (genes, drugs, diseases, variants,).  
Leverage these expansions to guide your searches and analysis.

# Deliverable
Return a concise summary only, following this strict format:
```
# Files saved:  
- filepath1: one-sentence description
- filepath2: one-sentence description  

Main findings:  
1-2 short sentences stating the key insight or next step.
```

Do not include detailed narratives, lists, or study summaries. Keep the entire output under 10 lines total.
"""
\end{lstlisting}

\subsection{DFRS agent prompt}\label{sup:dfrs_prompt}
\begin{lstlisting}
"""
You are a biomedical research assistant operating in an interactive coding environment with access to specialized tools, data sources, and analytical software.

# Objective
You are working under the directory `{workdir}`. Your goal is to perform depth-first exploration and analysis on the seed results provided (or identified), progressively refining hypotheses and extracting detailed insights.

# Workflow
1. Targeted Deep Search - Begin from the given seed results or initial query. For each, perform focused and detailed searches to gather in-depth related evidence, datasets, or contextual information.
2. Progressive Analysis - Use `code_exec_tool` to analyze each layer of information as you go deeper. Prioritize reasoning chains that appear most promising, and document intermediate findings.
3. Iterative Refinement - Based on analytical outcomes, determine the next layer or sub-topic to explore. Continue until you reach well-supported conclusions or no further meaningful depth is achievable.
4. Documentation & Synthesis - Summarize how each step of reasoning or exploration connects to prior layers. Record methodological notes, rationale for each branching path, and synthesized interpretations.
5. Save Outputs - Save the refined analyses, structured insights, and final synthesized results under `{workdir}`.


# Knowledge Base Integration
Before invoking search or executing code, use the knowledge base tools (`find_entities`, `find_related_entities`, `search_*`, `fetch_*_details`, etc.) to extract and expand biomedical entities (genes, drugs, diseases, variants).  
Leverage these expansions to guide your searches and analysis.

# Deliverable
Return a concise summary only, following this strict format:
```
# Files saved:  
- filepath1: one-sentence description
- filepath2: one-sentence description  

Main findings:  
1-2 short sentences stating the key insight or next step.
```

Do not include detailed narratives, lists, or study summaries. Keep the entire output under 10 lines total.
"""
\end{lstlisting}

\subsection{Evidence graph management prompt}\label{sup:evidence_graph_prompt}

\begin{lstlisting}
"""
# Memory Graph Protocol

## 1. What to Store
Keep only concise, high-value facts directly relevant to the research question.  
Each item in the graph should represent a unique and reusable concept - not a paraphrase.

### Biomedical Entities
- Type: GENE / PROTEIN (HGNC, NCBI Gene, UniProt, etc.)
- Type: DISEASE / PHENOTYPE (DOID, MeSH, UMLS)
- Type: CHEMICAL / DRUG (ChEBI, DrugBank, etc.)
- Type: CELL LINE / TISSUE (Cellosaurus, etc.)
- Type: PATHWAY / GENE_SET (Reactome, KEGG, GO, MSigDB, etc.)
- Type: PAPER (PMID; short title optional)
- Type: FINDING (only when they capture a concrete quantitative or mechanistic result)

### Relations
- Mechanistic: ACTIVATES, INHIBITS, BINDS, PHOSPHORYLATES, REGULATES_EXPRESSION
- Membership/annotation: MEMBER_OF_PATHWAY, HAS_GENESET_MEMBER, EXPRESSED_IN
- Association: ASSOCIATED_WITH, CO_OCCURS (use only when no precise predicate applies)
- Evidence-level: SUPPORTS, REFUTES, INCONCLUSIVE_FOR, CITES
- Provenance: DERIVED_FROM_KG:<name@version>

### Evidence Summaries
- Use short factual sentences (<= 30 words) capturing method, context, and numeric result if any.
- Context (species, cell type, assay) should **stay inside the observation**, not as separate context nodes unless reused by multiple findings.

## 2. Identifier & Naming Rules
- Prefer canonical CURIEs (HGNC:XXXX, DOID:XXXX, CHEBI:XXXX, PMID:XXXX).
- If canonical ID unavailable, keep the human-readable label and note its source KG.
- Paper entities: always start name with "PMID:"; optional short token after.
- Keep names <=5 words or <=40 characters.
- Never create multiple nodes for the same concept with case or wording variations.
- When encountering a near-duplicate:
  - If IDs match: update observations on the existing node.
  - If labels match (case-insensitive): treat as same node.
  - If labels differ but clearly same PMID or KG ID: merge; do not create new node.

## 3. Relation Standards
- Use the smallest consistent predicate set; do not introduce new verbs unless absolutely needed.
- Direction: always subject -> object, never reversed for stylistic reasons.
- Avoid generic ASSOCIATED_WITH edges when context is already captured in the observation text.
- Limit contextual edges:
  - Max two per finding (e.g., one to MEASURE, one to CELLTYPE).
  - Do not connect every assay/species as a separate ASSOCIATED_WITH edge.
- Each relation must reference at least one evidence source (PMID or KG provenance).

## 4. Graph Maintenance & Anti-Redundancy Rules
- GLOBAL graph is append-only but deduplicated by canonical ID and normalized label.
- Each merge cycle: <=10 new entities, <=16 new relations.
- Before creating any entity or relation, the agent must:
  1. Check if an equivalent already exists (by ID or normalized name).
  2. If found, update its observations instead of creating a new node.
- Conflicting results:
  - Keep both relations with distinct evidence; tag with `conflict_group:<id>`.
  - Do not duplicate entire entities just to hold alternative findings.
- Always prefer observations over new edges when adding simple context.
- Each paper appears exactly once (one node per PMID).
- Each finding appears exactly once per unique numeric or mechanistic result.
- Each context concept (species, cell type, assay) appears once per canonical ID.

## 5. Provenance & Review
- Every node and edge must include a provenance note (PMID or KG@version).
"""
\end{lstlisting}

\clearpage

\section{Supplementary figures}
% show detailed results of the trial development, study-by-study performance pairwise comparison

% show detailed results of the evidence gap discovery, study by study pairwise comparison

\begin{figure}[htbp]
    \centering
    \includegraphics[width=0.8\linewidth]{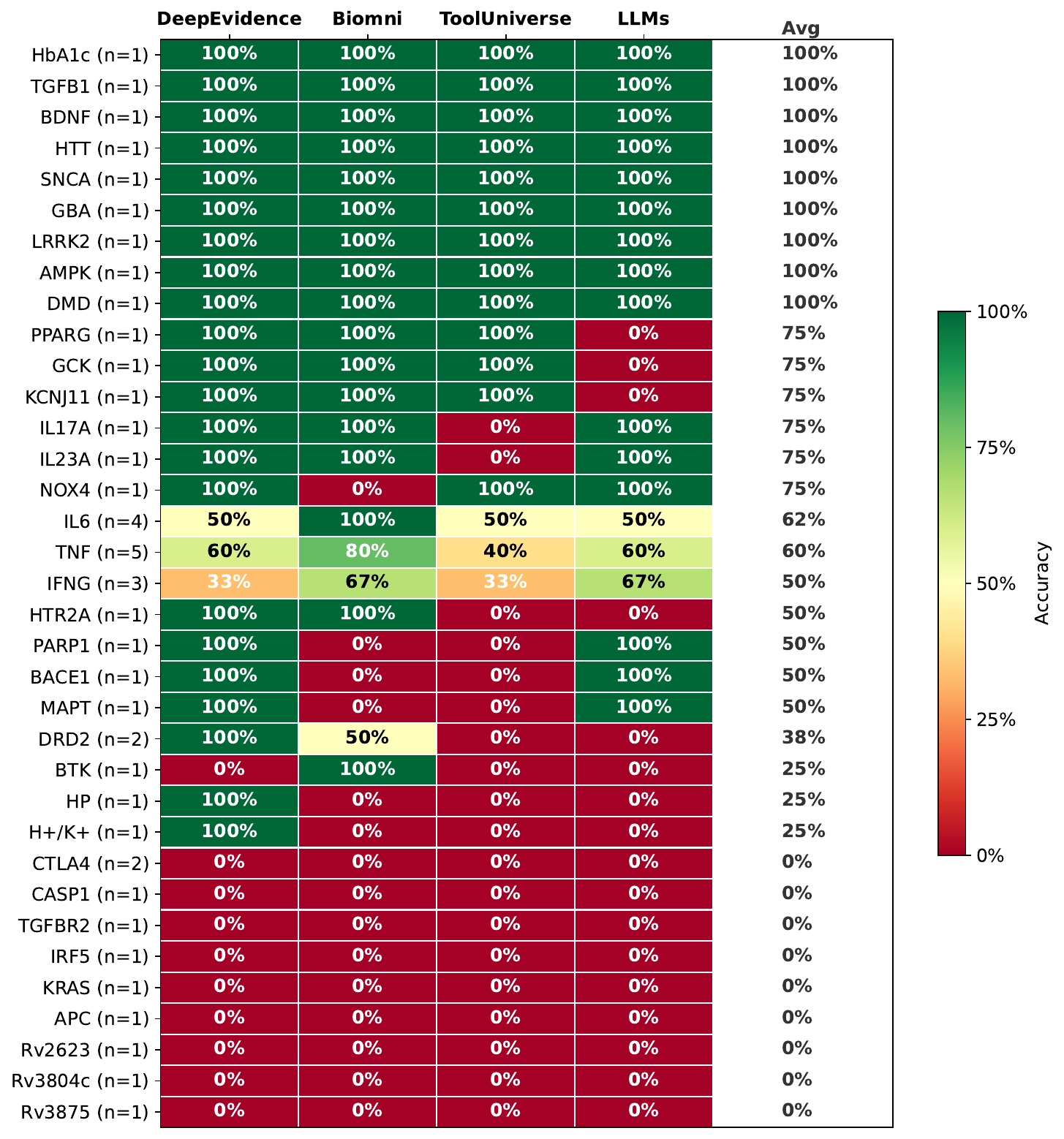}
    \caption{Gene-level correctness of four methods across evaluated targets in the target identification tasks.
The heatmap reports binary correctness (colored as 0\% or 100\%) for each method: DeepEvidence, Biomni, ToolUniverse, and LLMs, on individual gene targets. Rows correspond to gene targets (with the number of test instances per gene shown in parentheses), and columns correspond to methods; cell colors indicate whether the method produced a correct prediction for that gene (green) or not (red). The rightmost column shows the average correctness across methods for each gene. Overall, DeepEvidence achieves the broadest and most consistent gene coverage, Biomni shows moderate performance, and ToolUniverse and LLMs succeed on fewer targets. Results are descriptive, as each gene–method pair is evaluated once.}
    \label{fig:sup_target_identification_per_gene}
\end{figure}

\clearpage

\begin{figure}[htbp]
    \centering
    \includegraphics[width=\linewidth]{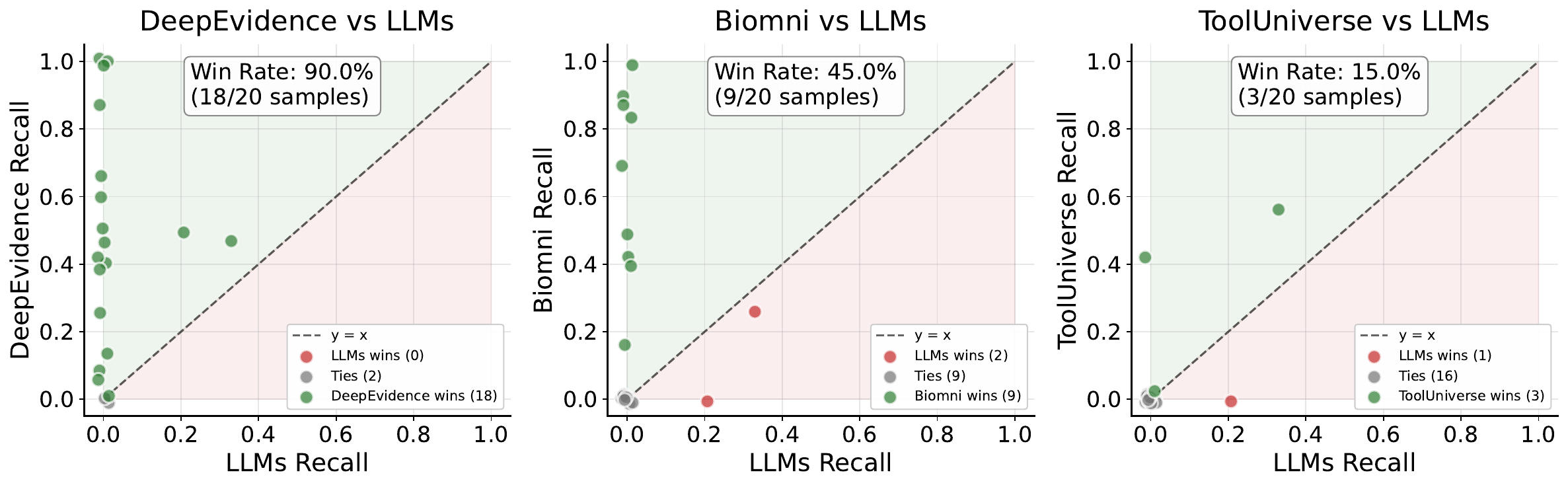}
    \caption{Pairwise recall comparison between agent-based systems and baseline LLMs in the evidence gap discovery tasks. Each panel plots task-level recall achieved by baseline LLMs (x-axis) against an agent-based system (y-axis): DeepEvidence (left), Biomni (middle), and ToolUniverse (right). Each point corresponds to one evaluation sample. The dashed diagonal indicates parity (y = x); points above the line denote cases where the agent outperforms the LLM, while points below indicate the opposite. Shaded regions highlight agent wins (green) and LLM wins (red). Insets report the win rate and counts across 20 reviews. DeepEvidence shows a substantial advantage over LLMs, whereas Biomni and ToolUniverse exhibit progressively weaker gains, highlighting differences in recall robustness across agent frameworks.}
    \label{fig:sup_evidence_gap_per_sample}
\end{figure}

\clearpage

\section{Supplementary tables}

\begingroup
\scriptsize   % or \footnotesize, \scriptsize (see notes below)

\setlength{\LTpre}{0pt}
\setlength{\LTpost}{0pt}
\begin{longtable}{p{3cm}p{5cm}p{7cm}}
\caption{Knowledge graph tools implemented in the \method framework.}
\label{tab:kg_tools} \\

\toprule
\textbf{Knowledge Base} & \textbf{Tool Name} & \textbf{Description} \\
\midrule
\endfirsthead

\multicolumn{3}{c}%
{{\tablename\ \thetable{} -- continued from previous page}} \\
\toprule
\textbf{Knowledge Base} & \textbf{Tool Name} & \textbf{Description} \\
\endhead

\midrule
\multicolumn{3}{r}{{Continued on next page}} \\
\endfoot

\bottomrule
\endlastfoot

BioThings (MyGene.info) & \texttt{search\_genes} & Search for genes by symbol name Entrez ID or Ensembl ID across multiple fields \\
 & \texttt{fetch\_gene\_details\_by\_ids} & Fetch detailed gene information including aliases pathways and GO terms for a list of gene IDs \\
\midrule
BioThings (MyChem.info) & \texttt{search\_drugs} & Search for drugs by name DrugBank ID ChEBI ID ChEMBL ID PubChem CID or InChI Key \\
 & \texttt{fetch\_drug\_details\_by\_ids} & Fetch detailed drug information including indications mechanism of action and pharmacology \\
\midrule
BioThings (MyDisease.info) & \texttt{search\_diseases} & Search for diseases by name MONDO ID DOID OMIM ID or MeSH ID \\
 & \texttt{fetch\_disease\_details\_by\_ids} & Fetch detailed disease information including phenotypes and cross-references \\
\midrule
BioThings (MyVariant.info) & \texttt{search\_variants} & Search for genetic variants by rsID gene HGVS notation chromosome position or ClinVar significance \\
 & \texttt{fetch\_variant\_details\_by\_ids} & Fetch detailed variant annotations including clinical significance and population frequencies \\
\midrule
ChEMBL & \texttt{search\_compounds} & Search ChEMBL database for compounds by name synonym or identifier \\
 & \texttt{get\_compound\_details} & Get detailed compound information including molecular properties structure and cross-references \\
 & \texttt{search\_similar\_compounds} & Find chemically similar compounds using Tanimoto similarity on SMILES structures \\
 & \texttt{search\_substructure} & Find compounds containing a specific chemical substructure using SMILES query \\
 & \texttt{batch\_compound\_lookup} & Process multiple ChEMBL IDs efficiently in batch mode \\
 & \texttt{search\_targets} & Search for biological targets (proteins protein complexes) by name type or organism \\
 & \texttt{get\_target\_details} & Get detailed target information including components synonyms and cross-references \\
 & \texttt{search\_by\_uniprot} & Find ChEMBL targets by UniProt accession number \\
 & \texttt{get\_target\_bioactivities} & Retrieve bioactivity measurements (IC50 Ki EC50) for a specific target \\
 & \texttt{get\_compounds\_for\_target} & Get active compounds for a target filtered by activity threshold \\
 & \texttt{get\_drug\_indications} & Search for therapeutic indications and disease areas for drugs \\
 & \texttt{get\_drug\_mechanisms} & Get mechanism of action and target interaction data for drugs \\
 & \texttt{get\_drug\_clinical\_data} & Get comprehensive clinical and drug development data including indications and mechanisms \\
 & \texttt{search\_drugs\_by\_indication} & Search for drugs treating a specific indication filtered by development phase \\
\midrule
ClinicalTrials.gov & \texttt{search\_trials} & Search clinical trials by condition intervention sponsor phase status or date range \\
 & \texttt{fetch\_trial\_details} & Fetch comprehensive trial information including eligibility criteria outcomes and study design \\
\midrule
Gene Ontology & \texttt{search\_go\_terms} & Search GO terms by name keyword or definition across molecular function biological process and cellular component ontologies \\
 & \texttt{get\_go\_term\_details} & Get detailed information for a GO term including definition synonyms and cross-references \\
 & \texttt{get\_go\_term\_hierarchy} & Get hierarchical relationships (ancestors descendants children) for a GO term \\
 & \texttt{validate\_go\_id} & Validate a GO identifier format and check if it exists in the ontology \\
 & \texttt{get\_ontology\_statistics} & Get statistics about GO ontologies including term counts and evidence codes \\
 & \texttt{get\_gene\_annotations} & Get GO annotations for a specific gene with optional taxonomy and evidence code filters \\
 & \texttt{get\_term\_annotations} & Get gene products annotated with a specific GO term \\
 & \texttt{get\_evidence\_codes} & Get list of GO evidence codes with descriptions and hierarchy \\
\midrule
Human Phenotype Ontology & \texttt{search\_hpo\_terms} & Search HPO terms by keyword HPO ID or synonym \\
 & \texttt{get\_hpo\_term\_details} & Get detailed information for an HPO term including definition synonyms and cross-references \\
 & \texttt{get\_term\_genes} & Get genes associated with a specific HPO phenotype term \\
 & \texttt{get\_term\_diseases} & Get diseases associated with a specific HPO phenotype term \\
 & \texttt{get\_gene\_phenotypes} & Get phenotypes associated with a specific gene \\
\midrule
KEGG & \texttt{search\_pathways} & Search biological pathways by name or keyword across all organisms \\
 & \texttt{get\_pathway\_info} & Get detailed pathway information including description participating genes and compounds \\
 & \texttt{get\_pathway\_genes} & Get list of genes involved in a specific KEGG pathway \\
 & \texttt{get\_pathway\_compounds} & Get list of compounds involved in a specific KEGG pathway \\
 & \texttt{search\_genes} & Search genes by symbol or description in KEGG GENES database \\
 & \texttt{get\_gene\_info} & Get detailed gene information from KEGG including pathways and orthologs \\
 & \texttt{search\_compounds} & Search chemical compounds in KEGG COMPOUND database \\
 & \texttt{get\_compound\_info} & Get detailed compound information including formula and reactions \\
 & \texttt{search\_diseases} & Search human diseases in KEGG DISEASE database \\
 & \texttt{get\_disease\_info} & Get detailed disease information including associated genes and drugs \\
 & \texttt{search\_drugs} & Search approved drugs in KEGG DRUG database \\
 & \texttt{get\_drug\_info} & Get detailed drug information including targets indications and interactions \\
 & \texttt{get\_drug\_interactions} & Retrieve drug-drug interaction information for multiple drugs \\
 & \texttt{search\_reactions} & Search biochemical reactions in KEGG REACTION database \\
 & \texttt{search\_enzymes} & Search enzymes by EC number or name \\
 & \texttt{search\_modules} & Search functional modules in KEGG MODULE database \\
 & \texttt{search\_ko\_entries} & Search KEGG Orthology entries by keyword \\
\midrule
NCBI Datasets & \texttt{search\_genes} & Search genes by symbol and taxonomy ID using NCBI Datasets API \\
 & \texttt{get\_gene\_info} & Fetch gene annotations genomic locations and orthologs \\
 & \texttt{search\_genomes} & Search genome assemblies by organism or accession \\
 & \texttt{get\_taxonomy\_info} & Retrieve taxonomic classification and lineage information \\
\midrule
OpenFDA & \texttt{search\_openfda\_drugs} & Search FDA-approved drugs by brand name generic name substance or application number \\
 & \texttt{search\_drug\_labels} & Search drug labeling for indications warnings dosage and prescribing information \\
\midrule
Open Genes & \texttt{search\_genes} & Search aging-related genes by symbol aging mechanism disease or protein class \\
 & \texttt{get\_gene\_by\_symbol} & Get detailed aging gene information including longevity evidence and confidence levels \\
 & \texttt{get\_calorie\_experiments} & Get caloric restriction experiment data related to aging \\
 & \texttt{get\_aging\_mechanisms} & Get list of aging mechanisms and associated genes \\
 & \texttt{get\_protein\_classes} & Get protein classes related to aging research \\
 & \texttt{get\_model\_organisms} & Get model organisms used in aging research \\
\midrule
Open Targets & \texttt{search\_targets} & Search therapeutic targets (genes) by symbol name or description \\
 & \texttt{get\_target\_details} & Get comprehensive target information including pathways tractability and cross-references \\
 & \texttt{get\_target\_associated\_diseases} & Get diseases associated with a target with evidence scores \\
 & \texttt{search\_diseases} & Search diseases by name synonym or EFO ID \\
 & \texttt{get\_disease\_details} & Get comprehensive disease information including therapeutic areas and hierarchy \\
 & \texttt{get\_disease\_associated\_targets} & Get targets associated with a disease with evidence scores \\
 & \texttt{get\_disease\_targets\_summary} & Get summary of all targets associated with a disease including top targets \\
 & \texttt{search\_drugs} & Search drugs by name or ChEMBL ID \\
 & \texttt{get\_drug\_details} & Get comprehensive drug information including type clinical phase and linked entities \\
 & \texttt{get\_target\_disease\_evidence} & Get detailed evidence linking a specific target-disease pair \\
 & \texttt{analyze\_association\_evidence} & Analyze the evidence types and sources for target-disease associations \\
\midrule
ProteinAtlas & \texttt{search\_proteins} & Search proteins by gene name symbol or description \\
 & \texttt{get\_protein\_info} & Get detailed protein information including expression data and subcellular location \\
 & \texttt{batch\_protein\_lookup} & Look up multiple proteins simultaneously by gene symbols \\
 & \texttt{get\_tissue\_expression} & Get tissue-specific RNA and protein expression levels \\
 & \texttt{get\_blood\_expression} & Get blood cell expression data for a protein \\
 & \texttt{get\_brain\_expression} & Get brain region expression data for a protein \\
 & \texttt{search\_by\_tissue} & Find proteins expressed in specific tissues \\
 & \texttt{get\_subcellular\_location} & Get subcellular localization annotations for a protein \\
 & \texttt{get\_pathology\_data} & Get cancer-related expression and survival data \\
 & \texttt{search\_cancer\_markers} & Search for cancer biomarker proteins \\
 & \texttt{get\_antibody\_info} & Get antibody validation information for a protein \\
\midrule
PubChem & \texttt{search\_compounds} & Search compounds by name CAS number SMILES InChI or formula \\
 & \texttt{get\_compound\_info} & Get detailed compound information including properties identifiers and synonyms \\
 & \texttt{get\_compound\_properties} & Get molecular properties for compounds \\
 & \texttt{get\_compound\_synonyms} & Get all synonyms for a compound \\
 & \texttt{search\_similar\_compounds} & Find structurally similar compounds using fingerprint similarity \\
 & \texttt{substructure\_search} & Find compounds containing a specific substructure \\
 & \texttt{get\_safety\_data} & Get safety and hazard information including GHS classifications \\
 & \texttt{get\_toxicity\_info} & Get toxicity information for a compound \\
 & \texttt{get\_assay\_info} & Get bioassay information for compounds \\
 & \texttt{get\_compound\_bioactivities} & Get bioactivity data across multiple assays \\
 & \texttt{get\_external\_references} & Get cross-references to external databases \\
 & \texttt{assess\_drug\_likeness} & Assess drug-likeness using Lipinski and other rules \\
\midrule
PubTator & \texttt{search\_papers} & Search PubMed articles using boolean queries with entity typing (@CHEMICAL @DISEASE @GENE) \\
 & \texttt{pubtator\_api\_fetch\_paper\_annotations} & Retrieve NER annotations (genes diseases chemicals variants) for PubMed articles \\
 & \texttt{pubtator\_api\_find\_entities} & Discover biomedical entities with autocomplete and standardized ID resolution \\
 & \texttt{pubtator\_api\_find\_related\_entities} & Find entities related through semantic predicates (TREAT CAUSE INTERACT INHIBIT ASSOCIATE) \\
\midrule
PubMed & \texttt{pubmed\_api\_search\_papers} & Search PubMed using E-utilities with boolean query support \\
 & \texttt{pubmed\_api\_get\_paper\_references} & Retrieve citation relationships (references and cited-by) for articles \\
 & \texttt{fetch\_paper\_content\_by\_pmid} & Fetch full-text content when available via PubMed Central \\
 & \texttt{get\_pubmed\_articles} & Get article metadata including title abstract journal and authors \\
 & \texttt{extract\_relevant\_sections} & Extract sections from paper content matching a pattern with context \\
\midrule
Reactome & \texttt{search\_pathways} & Search biological pathways by name or identifier in Reactome \\
 & \texttt{get\_pathway\_details} & Get detailed pathway information including hierarchy participants and reactions \\
 & \texttt{get\_pathway\_hierarchy} & Get parent and child pathways in the Reactome hierarchy \\
 & \texttt{get\_pathway\_reactions} & Get reactions that are part of a pathway \\
 & \texttt{get\_pathway\_participants} & Get molecules participating in a pathway \\
 & \texttt{find\_pathways\_by\_gene} & Find pathways containing a specific gene or protein \\
 & \texttt{get\_gene\_pathways\_dataframe} & Get pathways for a gene as a DataFrame \\
 & \texttt{get\_protein\_interactions} & Retrieve protein-protein interactions from pathway data \\
 & \texttt{find\_pathways\_by\_disease} & Find pathways associated with a disease \\
\midrule
UMLS & \texttt{search\_concepts} & Search UMLS Metathesaurus concepts by term or phrase \\
 & \texttt{get\_cui\_info} & Retrieve concept definitions semantic types and atoms for a CUI \\
 & \texttt{get\_atoms} & Get atoms (terms from source vocabularies) for a concept \\
 & \texttt{get\_definitions} & Get definitions for a concept from various sources \\
 & \texttt{get\_relations} & Retrieve semantic relationships between concepts \\
 & \texttt{get\_crosswalk} & Map identifiers across source vocabularies (ICD SNOMED MeSH RxNorm) \\
 & \texttt{get\_semantic\_type} & Get semantic type information from the UMLS Semantic Network \\
 & \texttt{get\_source\_concept} & Get concept information from a specific source vocabulary \\
\midrule
UniProt & \texttt{search\_proteins} & Search protein database by name gene organism function or localization \\
 & \texttt{get\_protein\_info} & Retrieve comprehensive protein information including sequence features and annotations \\
 & \texttt{search\_by\_gene} & Search proteins by gene name \\
 & \texttt{get\_protein\_sequence} & Get protein sequence in FASTA format \\
 & \texttt{get\_protein\_features} & Get protein features (domains active sites PTMs) \\
 & \texttt{get\_protein\_structure} & Get structural information and domain annotations \\
 & \texttt{get\_protein\_variants} & Get natural and disease-associated variants \\
 & \texttt{get\_protein\_pathways} & Get pathway associations from cross-references \\
 & \texttt{get\_protein\_interactions} & Retrieve protein-protein interaction data \\
 & \texttt{get\_protein\_homologs} & Find homologous proteins across species \\
 & \texttt{get\_protein\_orthologs} & Find orthologous proteins in different organisms \\
 & \texttt{search\_by\_function} & Search proteins by molecular function \\
 & \texttt{search\_by\_localization} & Search proteins by subcellular localization \\
 & \texttt{search\_by\_taxonomy} & Search proteins by taxonomic classification \\
 & \texttt{batch\_protein\_lookup} & Process multiple UniProt accessions efficiently \\
 & \texttt{get\_literature\_references} & Get literature references for a protein \\
 & \texttt{get\_external\_references} & Get cross-references to external databases \\
\midrule
Unified (Multi-source) & \texttt{search\_genes\_unified} & Search genes across BioThings KEGG and Open Targets with unified results \\
 & \texttt{fetch\_gene\_details\_unified} & Fetch comprehensive gene details from multiple sources with cross-referenced IDs \\
 & \texttt{search\_drugs\_unified} & Search drugs across BioThings OpenFDA KEGG Open Targets and ChEMBL \\
 & \texttt{fetch\_drug\_details\_unified} & Fetch comprehensive drug details aggregated from multiple authoritative sources \\
 & \texttt{search\_targets\_unified} & Search therapeutic targets across multiple databases \\
 & \texttt{fetch\_target\_details\_unified} & Fetch comprehensive target information with cross-database identifiers \\
 & \texttt{search\_pathways\_unified} & Search pathways across KEGG Reactome and other pathway databases \\
 & \texttt{fetch\_pathway\_details\_unified} & Fetch comprehensive pathway information from multiple sources \\
 & \texttt{search\_compounds\_unified} & Search compounds across PubChem ChEMBL and KEGG \\
 & \texttt{fetch\_compound\_details\_unified} & Fetch comprehensive compound information from multiple chemical databases \\

\end{longtable}
\endgroup

\begin{table*}
    \centering
    \caption{Example instance of Target Identification}
    \label{tab:target_identification_example}
    \begin{minipage}{0.95\textwidth}
        \centering
        \begin{tcolorbox}[title=Task Type: Target Identification]
            \textbf{Disease Context:} Ulcerative Colitis

            \vspace{1ex}
            \textbf{Question:}  
            Which genes represent the most effective therapeutic targets for modulating the inflammatory response in Ulcerative Colitis. Focus on targets that can be influenced to mitigate inflammation and improve patient outcomes. Select two most promising genes.

            \vspace{1ex}
            \textbf{Candidate Targets:}
            \begin{itemize}
                \item A. TNF : cytokine
                \item B. IL10 : cytokine
                \item C. VEGFA : growth factor
                \item D. MMP9 : enzyme
                \item E. PTPN2 : phosphatase
                \item F. SMAD7 : transcription regulator
                \item G. NOD2 : pattern recognition receptor
                \item H. IL6 : cytokine
                \item I. STAT3 : transcription factor
                \item J. CXCR2 : receptor
            \end{itemize}

            \vspace{1ex}
            \textbf{Correct Answer:}  
            A. TNF and H. IL6
        \end{tcolorbox}
    \end{minipage}
\end{table*}

\begin{table*}
    \centering
    \caption{Example instance of Mechanism of Action and Pathway Reasoning}
    \label{tab:mechanism_pathway_example}
    \begin{minipage}{0.95\textwidth}
        \centering
        \begin{tcolorbox}[title=Task Type: Mechanism of Action and Pathway Reasoning]
            \textbf{Question:}  
            Which experimental approaches can establish mechanistic linkage between molecule A and molecule B when both activate the same signaling pathway.

            \vspace{1ex}
            \textbf{Candidate Experimental Approaches:}
            \begin{itemize}
                \item A. Time resolved phosphorylation profiling to determine upstream downstream order.
                \item B. Genetic knockdown of molecule A to test whether B induced signaling is dependent on A.
                \item C. Co immunoprecipitation to test physical interaction between A and B.
                \item D. CRISPR knockout of molecule B to test whether A induced signaling is mediated through B.
                \item E. Reporter assays measuring pathway specific transcriptional activation under A or B perturbation.
                \item F. Using unrelated inhibitors to assure pathway independence.
            \end{itemize}

            \vspace{1ex}
            \textbf{Correct Answer:}  
            A, B, C, and D
        \end{tcolorbox}
    \end{minipage}
\end{table*}

\clearpage

\begin{table*}
    \centering
    \caption{Example instance of In Vivo Metabolic Flux Response Prediction}
    \label{tab:in_vivo_flux_example}
    \begin{minipage}{0.95\textwidth}
        \centering
        \begin{tcolorbox}[title=Task Type: In Vivo Metabolic Flux Response Prediction]
            \textbf{Question:}  
            In an investigation into the metabolic dependencies of lung cancer, which two mouse cohorts would likely show the most pronounced metabolic flux response upon inhibition of SHMT2.

            \vspace{1ex}
            \textbf{Candidate Mouse Cohorts:}
            \begin{itemize}
                \item A. Tumors with sustained reduction in tracer labeling
                \item B. Tumors exhibiting no change in metabolic markers
                \item C. Tumors with transient decrease in tracer uptake
                \item D. Tumors demonstrating consistent suppression of metabolic activity
                \item E. Tumors showing rapid recovery of metabolic function
                \item F. Tumors with maintained low levels of metabolic markers
                \item G. Tumors presenting a rapid increase in tracer labeling
            \end{itemize}

            \vspace{1ex}
            \textbf{Correct Answer:}  
            A and D
        \end{tcolorbox}
    \end{minipage}
\end{table*}

\begin{table*}
    \centering
    \caption{Example instance of Surrogate Endpoint Discovery}
    \label{tab:surrogate_endpoint_example}
    \begin{minipage}{0.95\textwidth}
        \centering
        \begin{tcolorbox}[title=Task Type: Surrogate Endpoint Discovery]
            \textbf{Drug:} Nerandomilast (D12975)

            \vspace{1ex}
            \textbf{Indication:} Idiopathic pulmonary fibrosis

            \vspace{1ex}
            \textbf{Question:}  
            Propose potentially plausible surrogate endpoint strategies for a novel Phosphodiesterase IV inhibitor in treating Idiopathic pulmonary fibrosis. Select biomarkers measured within 2 to 12 weeks that could serve as intermediate endpoints linking pharmacodynamic effects to clinical improvement in disease activity.

            \vspace{1ex}
            \textbf{Candidate Surrogate Endpoint Strategies:}
            \begin{itemize}
                \item A. Quantify cleaved caspase 3 positive cells in tumor biopsies at weeks 2 and 6; track longitudinally with clinical response assessment
                \item B. Analyze cytokine signaling pathway inhibition in peripheral blood cells at weeks 2 and 4; assess JAK STAT or NF kappa B pathway suppression
                \item C. Measure C reactive protein and erythrocyte sedimentation rate serially at weeks 2, 4, and 8; correlate with clinical disease activity scores
                \item D. Measure circulating biomarkers at weeks 2, 4, and 6 without tissue correlation
                \item E. Assess endothelial function via flow mediated dilation and arterial stiffness at weeks 4 and 8; correlate with vascular biomarkers
                \item F. Measure target occupancy using PET tracer at weeks 2, 4, and 8; assess direct binding as pharmacodynamic marker
                \item G. Quantify signaling protein activation at weeks 2, 4, and 8; track pathway modulation as molecular response marker
                \item H. Track circulating inflammatory cytokines (TNF alpha, IL 6, IL 1 beta) at baseline and weeks 1, 2, 4, 8; correlate with clinical response at week 12
                \item I. Perform tissue biopsy at week 8 only
            \end{itemize}

            \vspace{1ex}
            \textbf{Correct Answer:}  
            B, C, and H
        \end{tcolorbox}
    \end{minipage}
\end{table*}

\begin{table*}
    \centering
    \caption{Example instance of Sample Size Estimation}
    \label{tab:sample_size_estimation_example_app}
    \begin{minipage}{0.95\textwidth}
        \centering
        \begin{tcolorbox}[title=Task Type: Sample Size Estimation]
            \textbf{Condition:} Breast cancer female

            \vspace{1ex}
            \textbf{Study Arms:}
            \begin{itemize}
                \item \textbf{Arm A} Experimental Standard skin care supported by a reminder app
                \item \textbf{Arm B} Active comparator Standard skin care alone
            \end{itemize}

            \vspace{1ex}
            \textbf{Primary Outcome:} Prevention of grade greater than or equal to 2 radiation dermatitis assessed by CTCAE v5.0 through study completion average 4 weeks

            \vspace{1ex}
            \textbf{Question:}  
            Please estimate the sample size based on the assumption Chi square test two sided significance level of 5\% power of 80\% and a 2\% dropout rate

            \vspace{1ex}
            \textbf{Options:}
            \begin{itemize}
                \item A. 107
                \item B. 188
                \item C. 402
                \item D. 536
                \item E. 268
            \end{itemize}

            \vspace{1ex}
            \textbf{Correct Answer:} E. 268
        \end{tcolorbox}
    \end{minipage}
\end{table*}

\begin{table*}
    \centering
    \caption{Example instance of Drug Regimen Design}
    \label{tab:drug_regimen_design_example}
    \begin{minipage}{0.95\textwidth}
        \centering
        \begin{tcolorbox}[title=Task Type: Drug Regimen Design]
                     \textbf{Drug Combination:} Capecitabine, oxaliplatin, and irinotecan

            \vspace{1ex}
            \textbf{Population:} Locally advanced rectal cancer

            \vspace{1ex}
            \textbf{Question:}  
            You are planning a new early phase clinical trial to evaluate the combination of capecitabine, oxaliplatin, and irinotecan in locally advanced rectal cancer.
            The agents will be administered via oral and i.v.
            Based on this prior evidence, which trial design strategy would be most appropriate for this new combination trial.

            \vspace{1ex}
            \textbf{Options:}
            \begin{itemize}
                \item A. Starting dose near individual monotherapy MTD levels 75 to 90\% range for each agent Design De escalation design with pre specified dose reduction rules if DLTs observed DLT window standard DLT evaluation window Cycle 1 28 days Additional optional pharmacokinetic sampling standard Phase I DLT criteria Objective identify highest tolerable combination dose expecting doses near monotherapy levels
                \item B. Starting dose near individual monotherapy MTD levels 75 to 90\% range for each agent Design De escalation design with pre specified dose reduction rules if DLTs observed DLT window standard DLT evaluation window Cycle 1 21 days Additional optional pharmacokinetic sampling standard Phase I DLT criteria Objective identify highest tolerable combination dose expecting doses near monotherapy levels
                \item C. Starting dose conservative starting doses 30 to 50\% of individual monotherapy MTD range Design standard 3 plus 3 or model assisted escalation design BOIN or CRM DLT window standard to moderately extended DLT window Cycle 1 28 to 42 days Additional careful DLT criteria accounting for overlapping toxicities between agents Objective determine MTD and recommended Phase 2 dose for the combination
                \item D. Starting dose very conservative starting doses 15 to 25\% of monotherapy MTD range Design single agent lead in period followed by cautious combination escalation DLT window extended DLT evaluation window 2 cycles or 42 to 56 days Additional mandatory pharmacokinetic and drug drug interaction substudy intensive safety monitoring Objective establish feasibility and preliminary safety profile for this novel combination
                \item E. Starting dose approved or previously validated combination dose level Design limited dose adjustment primary focus on safety verification and tolerability DLT window standard DLT evaluation Cycle 1 21 to 28 days Additional expansion cohorts at target dose for additional safety and efficacy signals Objective verify safety and tolerability of established combination in new patient population
            \end{itemize}

            \vspace{1ex}
            \textbf{Correct Answer:} D
        \end{tcolorbox}
    \end{minipage}
\end{table*}

\begin{table*}
    \centering
    \caption{Example instance of Evidence Gap Discovery}
    \label{tab:evidence_gap_discovery_input}
    \begin{minipage}{0.95\textwidth}
        \centering
        \begin{tcolorbox}[title=Task Type: Evidence Gap Discovery]
            \textbf{Objectives:}  
            To assess the effects of audit and feedback on the practice of healthcare professionals and to examine factors that may explain variation in the effectiveness of audit and feedback.

            \vspace{1ex}
            \textbf{Selection Criteria:}  
            Randomised trials including cluster trials and cross over and factorial designs featuring audit and feedback defined as measurement of clinical performance over a specified period of time and provision of the resulting data to clinicians or clinical teams. Eligible trials reported objectively measured health professional practice outcomes.

            \vspace{1ex}
            \textbf{Previously Included Studies:}  
The prior review included the following studies (PubMed IDs):
 8129501, 16389536, 15592551, 10166596, 14695072, 12879828, 9431333, 16030275, 17365766, 9495403, 18415831, 2383413, 10600428, 10551192, 3230459, \ldots (omitted for avoiding clutter)
 
            \vspace{1ex}
            \textbf{Question:}  
            We are now updating this review to capture evidence published since the last search. Identify and list the PubMed IDs of new studies that should be included in the updated review.

            \vspace{1ex}
            \textbf{Target Answer Example:}  
            Newly identified PubMed IDs include  
            30994898, 20379742, 31209158, 9682686, 8520339, 15020170, 25670810, 16216030, 22727623, 22361211, \ldots (omitted for avoiding clutter)

            \vspace{1ex}
            \textbf{Note:}  
            Only the first ten PubMed IDs are shown here for brevity. The full answer contains many additional eligible studies omitted in this example.
        \end{tcolorbox}
    \end{minipage}
\end{table*}

%TC:endignore

\end{document}